\documentclass[preprint,11pt]{elsarticle}

\usepackage{graphicx}
\usepackage{import}
\usepackage{blindtext}
\usepackage{titlesec}
\usepackage{enumitem}
\usepackage[table,xcdraw]{xcolor}
\usepackage{tabularx}
\usepackage{booktabs}
\usepackage{ltablex}
\usepackage{todonotes}
\usepackage{hyperref}
\usepackage{booktabs}
\usepackage{lscape}
\usepackage{longtable}
\usepackage{cleveref}
\usepackage{listings}
\usepackage{multirow}
\usepackage{pgfplots}
\usepackage{hyperref}
\usepackage{dingbat}
\usepackage{floatrow}
\usepackage{capt-of}
\usepackage{caption}
\usepackage{longtable}
\usepackage[export]{adjustbox}
\usepackage{threeparttable}
\usepackage{caption}
\usepackage{subcaption}
\pgfplotsset{width=\textwidth,compat=1.8}
\usepackage{ulem}
\usepackage{soul}

\journal{Information and Software Technology}

\newcommand{\quickwordcount}{%
  \immediate\write18{texcount -1 -sum -merge \jobname.tex > \jobname-words.sum }%
  \input{\jobname-words.sum} words%
}

 \tikzstyle{mybox} = [draw=black, very thick, rectangle, rounded corners, inner ysep=5pt, inner xsep=5pt]
\usepackage{color}

\definecolor{darkmagenta}{rgb}{0.55, 0.0, 0.55}
\definecolor{darkgreen}{RGB}{6, 46, 3}
\definecolor{mygreen}{HTML}{1E8449}
\definecolor{amber}{rgb}{1.0, 0.49, 0.0}

\usepackage{color}
\newcommand\rf[1]{\textcolor{black}{#1}}
\newcommand\rs[1]{\textcolor{black}{#1}}
\newcommand\rt[1]{\textcolor{black}{#1}}
\newcommand\ra[1]{\textcolor{black}{#1}}
\newcommand\rb[1]{\textcolor{black}{#1}}

\newcommand\rd[1]{\textcolor{black}{#1}}

\begin{document}
\begin{frontmatter}
\title{Exploring Factors and Metrics to Select Open Source Software Components for Integration: An Empirical Study}

\author[TUNI]{Xiaozhou Li*}
\ead{xiaozhou.li@tuni.fi}

\author[TUNI]{Sergio Moreschini*}
\ead{sergio.moreschini@tuni.fi}

\author[TUNI]{Zheying Zhang}
\ead{zheying.zhang@tuni.fi}

\author[TUNI]{Davide Taibi}
\ead{davide.taibi@tuni.fi}

\address[TUNI]{Tampere University, Tampere (Finland) \\
$*$ the two authors equally contributed to the paper}

%
\begin{abstract}

 [Context] Open Source Software (OSS) is nowadays used and integrated in most of the commercial products. However, the selection of OSS projects for integration is not a simple process, mainly due to a of lack of clear selection models and lack of information from the OSS portals.  
 
[Objective] \rb{We investigate the factors and metrics that practitioners currently consider when selecting OSS. We also investigate the source of information and portals that can be used to assess the factors, as well as the possibility to automatically extract such information with APIs.} 

[Method] We elicited the factors and the metrics adopted to assess and compare OSS performing a survey among 23 experienced developers who often integrate OSS in the software they develop.   Moreover, we investigated the APIs of the portals adopted to assess OSS extracting information for the most starred 100K projects in GitHub. 

[Result]  
We identified a set consisting of 8 main factors and 74 sub-factors, together with 170 related metrics that companies can use to select OSS to be integrated in their software projects.  Unexpectedly, only a small part of the factors can be evaluated automatically, and out of 170 metrics, only 40 are available, of which only 22 returned information for all the 100K projects. \rt{Therefore, we recommend project maintainers and project repositories to pay attention to provide information for the project they are hosting, so as to increase the likelihood of being adopted}

[Conclusion]
OSS selection can be partially automated, by extracting the information needed for the selection from portal APIs. OSS producers can benefit from our results by checking if they are providing all the information commonly required by potential adopters. Developers can benefit from our results, using the list of factors we selected as a checklist during the selection of OSS, or using the APIs we developed to automatically extract the data from OSS projects. \\
\end{abstract}


\end{frontmatter}

\section{Introduction}

Open Source Software (OSS) has become mainstream in the software industry, and different OSS projects are now considered as good as closed source ones~\cite{RObles2019}\cite{Kilamo2020}. However, selecting a new OSS project requires special attention, and companies are still struggling to understand how to better select them~\cite{Lenarduzzi2020SEAA}\cite{LenarduzziOSS19}.

One of the main issues during the selection of OSS projects, is the lack of clear information provided by OSS providers \rt{about the software quality assessment}, and in particular the lack of automated tools that help the selection~\cite{Lenarduzzi2020SEAA}.

A local company hired our research group to ease and standardize the OSS selection process and to automate it as much as possible, to reduce the subjectivity and the effort needed for the evaluation phase. Currently, the company does not prescribe any selection model, and reported us that their developers commonly struggle to understand what they need to consider when comparing OSS projects.

In this paper, we investigate the first steps towards the definition of a semi-automated OSS evaluation model. 
Therefore, we extend our previous work~\cite{Lenarduzzi2020SEAA} by conducting a survey investigating the factors commonly considered by the companies when selecting OSS, the source of information that can be used to analyze these factors, and the availability of such information on the portals.

\rf{The goal of our work is to investigate and determine the factors that practitioners are currently considering when selecting OSS, to identify the sources (portals) that can be used to evaluate such factors mentioned by the practitioners, and to validate the public APIs that can be accessed to automatically evaluate those factors from the sources and portals.}

The research community has been studying OSS selection and evaluation from different perspectives. 

Researchers developed methods to evaluate, compare, and select OSS projects. Such methods and tools exploit different types of approaches, including manual extraction of data from OSS portals (e.g. OMM~\cite{OMM}, OpenBQR~\cite{Taibi2007BQR}, PSS-PAL~\cite{OSSPAL}). 

Researchers also proposed platforms for mining data from OSS repositories, that can also be used as the sources of information for the evaluation and comparison of OSS (e.g., The SourceForge Research Data Archive (SRDA)~\cite{madey2008sourceforge}, FLOSSmole~\cite{FLOSSMole}, FLOSSMetrics~\cite{Floss}, tools to provide dump of existing OSS portals (e.g. GHArchive~\cite{GHArchive}, GitTorrent~\cite{GHT}, ...), and tools to extract information from OSS portals (e.g. PyDriller~\cite{spadini2018pydriller}, CVSAnaly~\cite{robles2004remote}, OSSPesto~\cite{osspesto}...).  

Moreover, different approaches to evaluate software quality, often applied to OSS, have been proposed in research, including software metrics (e.g. Chidamber and Kemerer's metrics suite~\cite{CK}, Cyclomatic Complexity~\cite{McCabe}), tools to detect technical debt~\cite{Avgeriou2020} or to measure other quality aspects (e.g. Software Quality Index~\cite{googleQI}, Architectural Debt Index~\cite{Roveda2018}). 

Though the previous work provided a significant amount of results on OSS quality evaluation, OSS development data crawling, and OSS selection and adoption models, such works are still limited and not easy to apply in industry for selecting OSS because of various reasons: 

\begin{itemize}
    \item OSS selection models
    \begin{itemize}
        \item     Limited application in industry of the previous OSS selection models~\cite{Lenarduzzi2020SEAA}. The vast majority of models have never been adopted massively by industries with neither case studies nor success stories on the usage of these models therein. One of the potential reasons is that it is nearly not possible to have a generally accepted set of OSS selection criteria to use. The companies must adopt the criteria for their specific needs and constraints to achieve their business objectives. Another reason for such limited adoption of the selection models can be related to the lack of maturity of the models. The models lack clarity and guides about which metrics would offer the most relevant insights into the selection criteria.
    \end{itemize}
    \item OSS Mining Platforms such as  The SourceForge Research Data Archive (SRDA)~\cite{madey2008sourceforge}, GHArchive, GitTorrent~\cite{GHTorrent}
    \begin{itemize}
        \item They are designed for research purposes and are complex to use and often have different dependencies for developers that simply need to get data for an OSS project. As an example, GitArchive \cite{GHArchive} does not allow to directly \rf{query} the data with 
        an API, but needs to be accessed through Google Big Query \rf{ or dumping the files}. Moreover, the collection of all the information needed by the users to evaluate an OSS project requires to use several platforms. \rb{This study aims to provide an overview on what information is important to the companies and from what platforms to extract it when evaluating OSS projects. }
         
    \item They are often not maintained in the long term. As an example, The SourceForge Research Data Archive (SRDA)~\cite{madey2008sourceforge}  is not available anymore. The GHTorrent~\cite{GHTorrent} was created in 2013 with the last activity reported in 2019. More information on these platforms is available in Table~\ref{tab:rel-works-portals}.
    \end{itemize}
    \item Tools to evaluate software quality
    \begin{itemize}
        \item They are complex to use and often require effort for manual configuration and analysis on the target software.
        \item Most tools require expertise to understand which metrics should be used in which context, and how to interpret the evaluation results. They often provide an overload of information, but not always useful for every context. As an example, tools for assessing the quality and technical debt of software, such as SonarQube\footnote{SonarQube \url{http://www.sonarqube.org"}}, include more than 500 different rules to validate the source code, but only a limited amount of the rules that are commonly associated to specific qualities. 
        \item When existing tools focus on a specific set of quality metrics to do the evaluation, there is a lack of a tool that can aggregate the factors commonly considered during the selection of OSS.
        \item The existence of an OSS project community and of a health ecosystem is an informative indicator of the maturity of a software and its propensity for growth. Even though there are many community-related factors and metrics identified in the OSS selection model, some metrics can not be accessed directly from the project's repository and existing tools provide very limited support to analyze the data associated with the OSS project community and its support in OSS evaluation.
    \end{itemize}
    \item Existing Software Quality Models and Metrics such as the Architectural Quality index~\cite{Roveda2018}, but also metrics such as the Cyclomatic complexity~\cite{McCabe} or the presence of Code Smells~\cite{Fowler1999} or anti-patterns~\cite{brown1998antipatterns}. 
    \begin{itemize}
        \item Are usually targeting mainly on quality, while companies might be interested in other aspects while selecting OSS (e.g. Costs~\cite{DelBianco2010SERMA}, licenses, features, ...)
        
        \item Lack of comparison of the magnitude of the observed effect: different models return different outputs. As an example, previous works indicated that high levels of cyclomatic complexity might result in less readable source code. While besides that, the presence of some smells can be more harmful than others, but the analysis did not take into account the magnitude of such observed phenomenon. As an example, it is not clear if a piece of code with a cyclomatic complexity equals to 10 is twice more complex to read than the same piece of code with that of 5. The same applies to the comparison of several other metrics, including the presence of different amount of code smells.  Thus, the comparison of the results of the metrics, increases the complexity of the analysis and comparison between projects. 
        \item Lack of complex and historical analysis. A complete comparison of an OSS might require not only the analysis of the latest snapshots of a project, but a historical analysis, thus increasing the complexity and the effort required to perform an analysis. 

    \end{itemize}
 \end{itemize}   
    In addition, lack of expertise in companies, in particular on software quality, hinders practitioners to select the most suitable quality models for comparing the projects. 

To cope with the aforementioned issues, this paper aims at corroborating and extending previous empirical research on OSS selection and adoption, so as to enable, not only our target company to assess and compare OSS but also other companies. 
More specifically, this study aims at extending our previous work~\cite{Lenarduzzi2020SEAA} 
from different point of views:
\begin{itemize}
    \item  We performed a survey to update the \rb{common} factors that are currently considering when selecting OSS and we compared them with the factors considered in the past (elicited in the Euromicro/SEAA SLR~\cite{Lenarduzzi2020SEAA})
\item We analyzed the source of information and portals that can be used to assess the aforementioned factors
\item We analyzed the public APIs that can be accessed to automatically assess the factors from the aforementioned
portals
\item We extracted the information for 100K projects, to validate their availability.
\end{itemize}
\rb{The source of information associated with the  factors and metrics adopted to measure them will help developers to understand and adopt OSS selection models for their specific needs and constraints. Moreover, they will help to remind OSS producers to provide information commonly expected by the potential users of the software.}

\rb{ Together with the validated APIs to automatically extract the assessment information, the result of the work forms a critical step towards developing semi-automatic tools to facilitate the practice of OSS selection.}



The remainder of this paper is structured as follows.
Section~2 presents related works. Section~3 describes the research method we adopted to achieve our goals. Section~4 reports the results while Section~5 discuss them. Section~6 finally draws conclusions and future works. 
\label{sec:Introduction}

\section{Related Work}


To cope with the need of selecting valuable OSS projects, several evaluation models have been proposed  (e.g. \cite{Duijnhouwer_2003}, \cite{Golden_2008}, \cite{Davide_etal_2007} and \cite{Semeteys_2008}). 
At the same time, different research groups proposed project aggregators to ease the access of  different information on OSS, measures and other information. 
Last, but not least, research in mining software repositories also evolved in parallel, and different researchers provided datasets of OSS projects,  portals and tools to extract information from OSS projects.

In the remainder of this Section, we summarize related work on the factors adopted to evaluate OSS,  OSS evaluation and selection models, OSS aggregator portals, tools for OSS repository mining and tools for OSS analysis and audit. 

\subsection{The factors considered during the adoption of OSS}

In \rt{the} systematic literature review \rs{(SLR)} on OSS selection and adoption models~\cite{Lenarduzzi2020SEAA} \rf{that we are extending in this work}, we analyzed
60 empirical studies, including 20 surveys, 5 lessons learned on OSS adoption motivation and 35 OSS evaluation models. 

Regarding the common factors of OSS selection and adoption, eight main categories were reported by the selected studies, including, \textit{Community and Adoption}, \textit{Development
process}, \textit{Economic}, \textit{Functionality}, \textit{License}, \textit{Operational
software characteristics}, \textit{Quality}, \textit{Support and Service}. For each category, sub-factors or metrics are reported. Results show that not all factors were considered equally important according to evaluation models and to surveys and lessons learned. For example, factor \textit{cost} is considered much more important by the surveys than by the models when, on the contrary, the importance of factor \textit{maturity} is seen oppositely. Furthermore, certain factors are considered important by both groups, such as, \textit{Support and Service}, \textit{Code Quality}, \textit{Reliability}, etc.

Table~\ref{tab:rel-works-portals} lists sources of information mentioned in the related works to assess the common factors considered during the adoption of OSS. The table only reports the indirect sources of information. Direct sources of information such as the official portal, or the versioning system (e.g. GitHub, GitLab) are not mentioned in the table..

\subsection{OSS Evaluation and Selection Models}

Within the 35  OSS models identified in our previous literature review~\cite{Lenarduzzi2020SEAA}, 21 (60\%) were built via case study, with 5 via interview, 5 via experience and the other 4 via the combination of interview and case study. All proposed models provide either checklist (13 models) or measurement (8 models) or both (14 models) as their working approaches. On the other hand, regarding the studies with surveys and lesson learned, the majority (13) target at adoption motivation identification with the remainders on other scopes. Furthermore, 12 tools are introduced in 22 of the given studies; however, only two out of the 12 are properly maintained.

All the models propose to evaluate OSS with a similar approach: 
\begin{itemize}
    \item \textit{Identification of OSS candidates}. In this step, companies need to identify a set of possible candidates based on their needs. 
    \item \textit{Factors evaluation} A list of factors are then assessed, by extracting the information or measures from the OSS portals, or by measuring/running the project candidates
    \item \textit{Project scoring} The final score is then normalized based on the importance of each factor and the final evaluation is computed.
\end{itemize}

The Open Source Maturity Model (OSMM), was the first model proposed~\cite{Duijnhouwer_2003}\cite{Golden_2008} in the literature. OSMM is an open standard that aims at facilitating the evaluation and adoption of OSS. The evaluation is based on the assumption that the overall quality of the software is proportional to its maturity.
The evaluation is performed in three steps:
\begin{enumerate}
\item	Evaluation of the maturity of each aspect. The considered aspects are: the software product, the documentation, the support and training provided, the integration, the availability of professional services. 
\item	Every aspect is weighted for importance. The default is: 4 for software, 2 for the documentation, 1 for the other factors.
\item	The overall maturity index is computed as the weighted sum of the aspects’ maturity.
\end{enumerate}
The OSMM has the advantage of being simple. It allows fast (subjective) evaluations. However, the simplicity of the approach is also a limit: several potentially important characteristics of the products are not considered. For instance, one could be interested in the availability of professional services and training, in details of the license, etc. All these factors have to be ‘squeezed’ into the five aspects defined in the model.

The Open Business Readiness Rating (OpenBRR)~\cite{Wasserman_etal_2006} is an OSS evaluation method aiming at providing software professionals with an index applicable to all the current OSS development initiatives, reflecting the points of view of large organizations, SMEs, universities, private users, etc. 
The OpenBRR is a relevant step forward with respect to the OSMM, since it includes more indicators, the idea of the target usage, and the possibility to customize evaluations performed by other, just by providing personalized weights. With respect to the latter characteristics, the OpenBRR has however some limits: one is that for many products it is difficult to choose a “reference application” that reflects the needs of the users; another is that there are lots of possible target usages, each with its own requirements; finally, the evaluation performed by a user could be not applicable to other users. In any case, the final score is a synthetic indicator to represent the complex set of qualities of a software product.
On the official OpenBRR site several evaluations were available, and originally provided as spreadsheet. However the OpenBRR website and tools are not available anymore.

The Qualification and Selection of Open Source Software (QSOS)~\cite{Semeteys_2008} works similarly as OpenBRR, but requires first to create an Identity Card (IC) of each project, reporting general information (name of the product, release date, type of application, description, type of license, project URL, compatible OS, …), then to evaluate the available services, functional and technical specifications and grade them (in the 0..2 range). Then, evaluators can specify the importance of the criteria and their constraints. Finally a normalized score is computed to compare the selected project candidates.  
Although the method is effectively applicable to most OSS, the QSOS approach does not represent a relevant step forward with respect to other evaluation methods. Its main contribution is the set of characteristics explicitly stated which compose the IC, and the provision of a guideline for the consistent evaluation of these characteristics. The evaluation procedure is rigid. For instance, it requires to define the IC of each OSS under evaluation, even if they are not completely matching the requirements. Such a procedure is justified when the ICs of products are available from the OS community before a user begins the evaluation. However even in this case it may happen that the user needs to consider aspects not included in the IC: this greatly decreases the utility of ready-to-use ICs. The strict guidelines for the evaluation of the IC, necessary to make other users’ scoring reusable, can be ill suited for a specific product or user. Finally, even though in the selection criteria it is possible to classify requirements as needed or optional, there is no proper weighting of features with respect to the intended usage of the software.

OpenBQR~\cite{Taibi2007BQR} works in a similar way as OpenBRR, but requires the evaluators to first specify the importance of the factors, and then to assess the projects, so as to avoid to invest time evaluating factors that are not relevant for the specific context. 
OpenBQR is an important step forward in terms of effort required to evaluate the projects. 

OSS-PAL~\cite{OSSPAL} works similarly as QSOS, but proposed to introduce a semi-automated evaluation, supported by an online portal. Unfortunately, the portal seems to be only a research prototype, and does not collect any data automatically.

All the aforementioned models have some drawbacks:
\begin{itemize}
    \item	Existing methods usually focus on specific aspects of OSS. \rf{For example, the OSMM focuses on software maturity, but misses some potentially interesting characteristics like license compliance or security for the quality assessment. On the other hand, methods like OpenBRR, QSOS, OpenBQR, etc. provide a set of indicators reflecting a wide range of potential users’ viewpoints for the quality assessment~\cite{DelBianco2010IFIP}. This requires individuals to identify the importance of assessment factors according to their needs and introduces extra effort and complexity to adopt a method for practice.}
    \item 	\rf{The OSS evaluation requires effort to run the software and to extract information from the OSS portals. The assessment process of existing methods is not optimized. }Methods \rf{such as QSOS }proceed to evaluate indicators before they are weighted, so some factors may be measured or assessed even if they are later given a very low weight or even a null one. This results in unnecessary waste of time and effort.
    \item The dependence of the users of OSS is not adequately assessed, especially the availability of support over time and the cost of proprietary modules developed by third parties.
    
\end{itemize}

Besides the aforementioned models, other approaches to evaluate the trustworthiness~\cite{DelBianco2011}\cite{DelBianco2010IFIP}\cite{Lavazza2010ACMIEEE}\cite{Taibi2008IFIP}, the reliability~\cite{Lavazza2012ACMSYMP}\cite{Tosi2017ECISME} of OSS, and the quality of the portals~\cite{Lavazza2012LNBIP}\cite{Basilico2011} of OSS have been proposed.

\subsection{OSS Aggregators}

Many platforms have been developed to collect and share OSS-related data, enabling a quick extraction of the information on different OSS projects. 

Ohloh was one of the first project aggregators~\cite{Bruntink14}\cite{allen2009ohloh} on the market (2004) aimed at indexing several projects from different platforms (GitHub, SurceForge, ...). 
In 2009, Ohloh was acquired by Geeknet, owners of  SourceForge~\cite{SourgeForge} that then sold it to Black Duck Software in 2010. Black Duck, was already developing a product for OSS audit, with a particular focus on the analysis of the license compatibility, and integrated Ohloh's functionality with their products. In 2014, Ohloh became ''Black Duck Open Hub''~\cite{OpenHub}. Finally,  Synopsys acquired Black Duck and renamed the  Black Duck Open Hub into ''Synopsys Open Hub''. 
Synopsis Open Hub is currently the only continuously updated OSS aggregator that include information of different OSS projects from different sources (versioning control systems, issue tracking systems, vulnerability databases). On January 2021, OpenHub indexed nearly 500K projects, and more than 30 billions of lines of code. 
It provides flexibility for users to select the metrics to compare project statistics, languages, repositories, etc. However, it lacks the OSS evaluation facilities that allow to adjust the importance of selected metrics according to users’ needs for automatically scoring the candidate software. In addition, it lacks information related to the community popularity, documentation, availability of questions and answers and other information. 

Other OSS aggregators have been proposed so far. FlossHub~\cite{FlossHub} and FLOSSMole~\cite{FLOSSMole} had similar goal of OpenHub. However, they have not been updated in the last years. 
FlossMetrics~\cite{Floss} had the goal of providing software metrics on a set of OSS. However, it has also been abandoned in 2010. 

The Software Heritage~\cite{cosmo_Zacchiroli_2017}, differently than the previously mentioned platforms,  has the goal of collecting and preserving the history of software projects, and is not meant to enable the comparison  or to provide support for selecting OSS.  The project is sponsored by different companies and foundations, including the UNESCO foundation.  The Software Heritage could be used as a source of information to analyze the activity of a project. However, its access is not immediate, and users need to use APIs to get detailed data on the projects. 

Other platforms, designed for supporting mining software activities, might also be used for obtaining relevant information from OSS. 
In particular, the Sourceforge research data archives \cite{madey2008sourceforge} shared the SourceForge.net data with academic researchers studying the OSS software phenomenon; GH Archive \cite{GHArchive} records the public GitHub timeline and makes it accessible for further analysis; and the GHTorrent project \cite{Gousios2013} creates a mirror of data offered through the Github REST API  to monitor the event timeline, the event contents, and the dependencies. 

\subsection{OSS Repository mining tools}

\rt{Besides the platforms that aggregate heterogeneous metric providers to track repositories associated with a wide range of OSS projects, there are also research prototypes or projects} to mine information from given repositories. In particular, BOA \cite{Dyer_etal_2013,Dyer_etal_2015} provide support to mine source code and development history from project repositories using the domain-specific language. 
Candoia~\cite{Candoia} also provided a platform for mining and sharing information from OSS projects.  
RepoGrams~\cite{Rozenberg2016}\cite{RepoGrams} allows to visually compare projects based on the history of the activity of their git repositories.

Other groups developed tools not aimed at supporting the selection of OSS, but that can be used as valuable sources of information. As an example,  PyDriller~\cite{spadini2018pydriller} can be used to obtain detailed information from commits. 

\rf{Surprisingly, none of the previously mentioned papers cited other tools such as Cauldron or SourceCred. Cauldron\footnote{\textbf{Cauldron:} https://cauldron.io} is a free open source software that is used to collect information from multiple sources as different information are retrieved. SourceCred\footnote{\textbf{SourceCred:} https://sourcecred.io/docs/} is an OSS technology which analyzes a project and determines the contributions of individuals in it. It is built on the idea that communities matters but also that the work of singles need to be visible and rewardables.} 

\rf{
The Community Health Analytics Open Source Software (CHAOSS)\footnote{\textbf{CHAOSS project:} https://chaoss.community/about/} project. CHAOSS, a Linux Foundation project, also developed tools to measure OSS projects, and in particular to measure community health, to  analyze software community development and to develop programs for the deployment of metrics not attainable through online trace data.}

\rf{Different  European projects also developed tools for mining data from OSS repositories. The EU H2020 CROSSMINER project\footnote{\textbf{CROSSMINER project:} https://www.crossminer.org} \cite{dirocco2021} includes techniques and tools for extracting knowledge from existing open source components generating relevant recommendations for the development of user's projects. The recommendation system focuses on 4 main activities:}
\rf{\begin{itemize}
    \item \textbf{Data Preprocessing:} containing tools to extract metadata from repositories
    \item \textbf{Capturing Context:} uses metadata to generate knowledge for mining functionalities
    \item \textbf{Producing Recommendation:} IDE to generate recommendations.
    \item \textbf{Presenting Recommendation:} IDE to show recommendations.
\end{itemize}
}
\rf{
QUALOSS (Quality in Open Source Software)~\cite{QUALOSS} and QualiSPo (Quality Platform for Open Source Software)~\cite{DelBianco2010QUaliSPO}~\cite{DelBianco2011} projects aimed at identifying quality models to evaluate the quality and the trustworthiness of OSS. Both projects proposed different tools for extracting data from repositories, to calculate software metrics and to identify possible issues in the code or in the community activity. However, none of the tools developed is currently active, and several of them are not  available anymore (e.g., QualiSPo~\cite{DelBianco2010QUaliSPO})}

\subsection{OSS audit and analysis tools}

Companies like Synopsys (formerly BlackDuck) \cite{synopsys} and 
WhiteSource \cite{WhiteSource} provide  solutions to software composition analysis and offer services of  the assessment of OSS quality and code security. 
Synopsis focuses on their professional services of the license compatibility while WhiteSource emphasizes the open source management to offer services such as viewing the state of OSS components, their license compliance, and the dependencies; prioritizing components' vulnerabilities based on how the proprietary code is utilizing them; analyzing the impact of the vulnerabilities, etc. 

Different tools to assess specific qualities are also available on the market. As an example, companies can use tools such as SonarQube~\cite{SonarCloud} or Sonatype~\cite{sonatype} to evaluate different code-related qualities such as the standard compliance or the technical debt. Or Security-specific tools such a WhiteHat Security~\cite{Whitehat}, Kiuwan~\cite{kiuwan} or others to evaluate the security vulnerabilities. 


\subsection{Gaps of the current OSS assessment models}
The different OSS assessment models, tools and platforms provide a possibility to assess OSS projects mainly from the perspectives of license obligation, application security, code quality, etc. They are about the state of software and its quality and comprise an essential part of the assessment model for OSS selection and adoption \cite{Sbai2018}\cite{Lenarduzzi2020SEAA}. Besides, activities, supports, or other projects surrounding a project form an important perspective demonstrating if a project exists in a lively ecosystem \cite{JANSEN2014}. In particular, metrics such as response times in Q\&A forums and bug trackers, the active contributors and their satisfaction, the user’s usage and their satisfaction, the number of downloads, the number of forks, bug-fix time, etc. are informative references indicating the productivity and a propensity for growth of the OSS project community. Some of the measures can be cross-referenced from different data sources, while some need further analysis based on the collected data. To the best of our knowledge, no portal has effectively taken these community-related factors into account when providing service to evaluate and compare OSS projects.

Furthermore, companies have their distinct strategies, needs, and constraints to adopt OSS projects in software development \cite{Lenarduzzi2020SEAA}. \rt{After practitioners identify a list of candidates that cover the expected features, meet requirements, and fit with the existing technical solution, }they specify the importance of the selection criteria \rt{, complying with the company's needs and restrictions}. As highlighted in our previous systematic literature review~\cite{Lenarduzzi2020SEAA}, it is impractical for companies to study every software assessment model to select the one that fits their needs best. 
Therefore, there is a need to call for the OSS evaluation and selection tools that not only guides developers to adapt OSS assessment criteria by identifying and weighing the ones fitting in a specific scenario, but also automates the process of assessing and comparing among a set of selected software based on information which can be extracted from the public APIs of available portals.


\begin{table}[ht!]
\caption{OSS Source of information and portals reported in the literature} 
\label{tab:rel-works-portals}
\scriptsize
\begin{tabular}{|l|p{1.1cm}|p{1.2cm}|l|} \hline

\textbf{Portal}	& 	\textbf{Created on} 	&	\textbf{Last  \quad  activity}	& 	\textbf{Information Stored}	\\ \hline
\multicolumn{3}{|l}{\textbf{	OSS Aggregators}	}			&		\\ \hline
\cite{SoftwareHeritage}	Software Heritage	& 	2018	&	2020	& 	Source Code	\\
\cite{OpenHub}	OpenHub	& 	2004	&	2020	& 	OSS tracker	\\
\cite{FlossHub}	FlossHub	& 	2008	&	2018	& 	Research portal	\\
\cite{SourgeForge}	SourceForge Research Data	& 	2005	&	2008	& 	Statistics	\\
\cite{GHArchive}	GH Archive	& 	2012	&	2020	& 	Timeline Record	\\
\cite{GHT}	GH Torrent	& 	2013	&	2019 & 	GH event monitoring	\\
\cite{FLOSSMole}	FLOSSMole	& 	2004	&	2017	& 	Project data	\\
\cite{PROMISE}	PROMISE	& 	2005	&	2006	& 	Donated SE data	\\
\cite{Floss}	FLOSSMetrics	& 	2006	&	2010	& 	Metrics and Benchmarking	\\ \hline
\multicolumn{3}{|l}{	\textbf{Audit and Analysis Tools}	}			&		\\ \hline
\cite{WhiteSource}	WhiteSource	& 	2011	&	2020	& 	Security	\\
\cite{FossID}	FossID	& 	2016	&	2020	& 	OS Compliance and Security		\\ 
\cite{synopsys}	Synopsys (formerly BlackDuck)	& 	2012	&	2020	& 	legal, security, and quality risks	\\
\cite{SonarCloud}	SonarQube	& 	2006	&	2020	& 	Code quality and security	\\
\cite{Whitehat}	WhiteHat	& 	2001	&	2020	& 	Software composition analysis	\\
\cite{SonarCloud}	SonarCloud	& 	2008	&	2020	& 	Software quality analysis	\\ \hline
\multicolumn{3}{|l}{	\textbf{OSS Mining Data Tools}	}			&		\\ \hline
\cite{BOA}	BOA	& 	2015	&	2019	& 	Source code mining	\\
\cite{Candoia}	Candoia	& 	2016	&	2017	& 	Software repository mining		\\ 
\cite{RepoGrams}	RepoGrams	& 	2016	&	2020	& 	OSS Comparison	\\ \hline

\multicolumn{3}{|l}{	\textbf{Questions and Answers portal}	}			&		\\ \hline
\cite{StackExchange}	Stack Exchange	& 	2009	&	2020	& 	Q\&A	\\ 
\cite{reddit}	Reddit	& 	2009	&	2020	& 	Q\&A	\\ \hline
\end{tabular}
\end{table}

\label{sec:related-works}


\section{Research Method}
\subsection{Goal and Research Questions}

Our goal is to investigate and determine the factors that practitioners are currently considering when selecting OSS, to identify the sources (portals) that can be used to evaluate such factors mentioned by the practitioners, and to validate the public APIs that can be accessed to automatically evaluate those factors from the sources and portals.

To achieve the aforementioned goals, we defined four main research questions(RQs).
\begin{itemize}
    \item [\textbf{RQ1.}] What factors are practitioners considering when selecting OSS projects to be integrated in the software they develop? \\
    In this RQ we aim at collecting the information adopted by practitioners when selecting projects to be integrated in the software they develop. We are not considering OSS products \rt{supporting} \rt{software development process and the management} such as IDEs, Office Suites, but software libraries, frameworks or any other tool that will be integrated and packaged as part of the product developed by the company. 
    
    \item [\textbf{RQ2}] Which metrics are used by practitioners to evaluate the factors adopted during the selection of OSS? \\
    In this RQ we aim at identifying the metrics adopted by practitioners to evaluate the factors they are interested to assess. As an example, practitioners might assess the size of the community checking the number of committer in the repository, or might check the size of the project by checking the number of commits in the repository or even downloading the software and measuring its size in lines of code. 
    \item [\textbf{RQ3}]  Which source of information and portals are used to assess OSS?\\
    In this RQ we aim at understanding which portals or other sources are used by practitioners to evaluate the  factors identified in RQ1, based on the metrics reported in RQ2. 
    
    \item [\textbf{RQ4}] Which factor can be extracted automatically from OSS portals?\\
    In this RQ, we aim to systematically analyze the common portals hosting OSS, to identify the information that can be extracted via APIs.  
 \end{itemize}   
    In order to answer our RQs, we conducted our work in three main steps:
    \begin{itemize}
        \item [Step 1:] Interviews among experienced software developers and project managers to elicit the factors affecting the OSS selection (RQ1), the metrics (RQ2) and the sources of  information 
        they adopt (RQ3).
        \item [Step 2:] Analysis of the APIs of the source of information (portals) identified in RQ3. 
        
        \item [Step 3:] Analysis of the availability of the metrics collected in the previous step (RQ2) in the public API of the sources of information adopted by practitioners (RQ3) among 100k projects (RQ4). 
    \end{itemize}

\begin{figure}[ht!]
\centering\includegraphics[width=\linewidth]{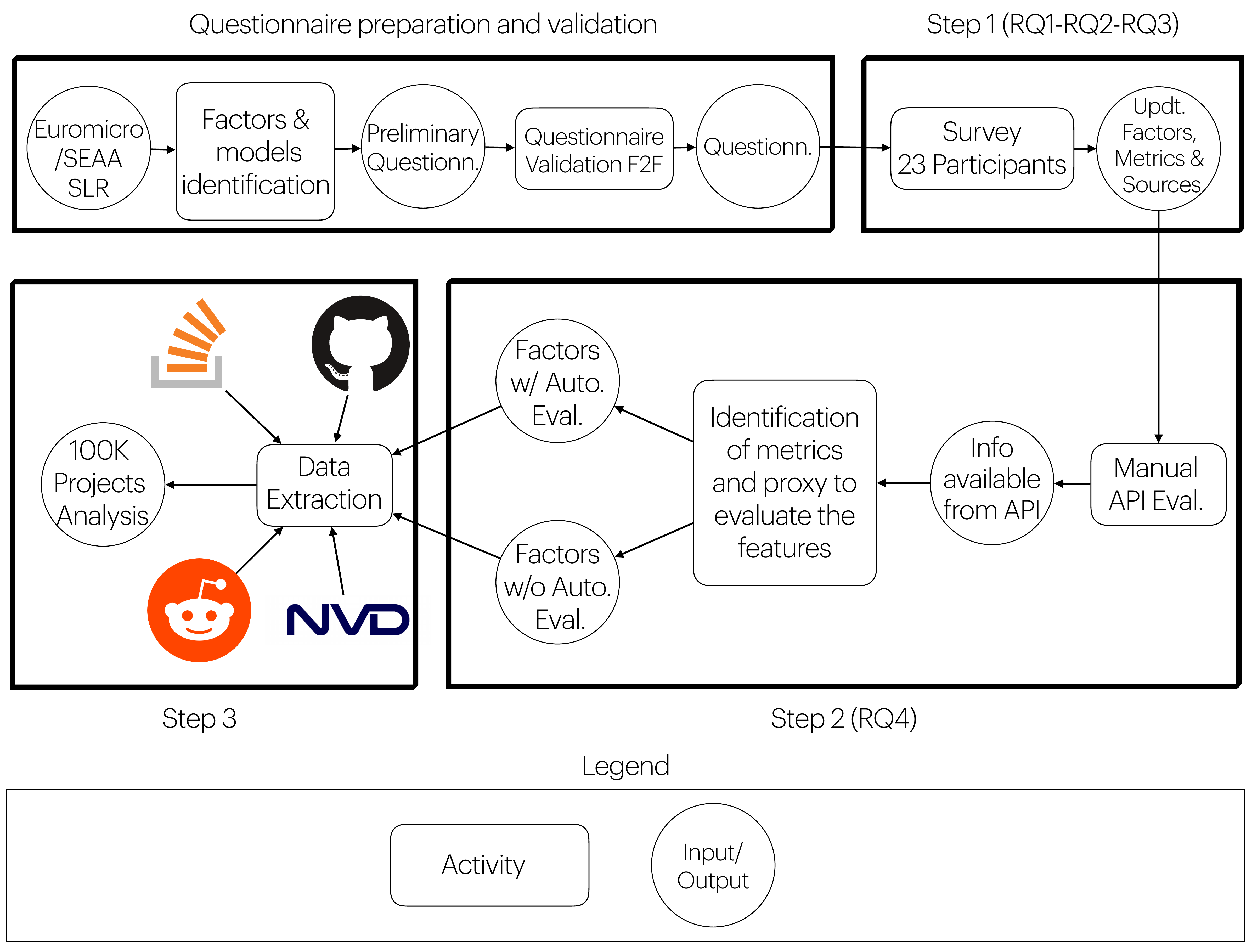}
\caption{The Study Process}
\label{fig:process}
\end{figure}

Figure~\ref{fig:process} depicts the process adopted in this work. The detailed process is reported in the remainder of this section.

\subsection{Step 1: Interviews on the factors considered when selecting OSS}
In order to elicit the factors adopted by practitioners when selecting an OSS in the software product development process, we designed and conducted a semi-structured interview based on a questionnaire.

\rd{\subsubsection{The Interview Population}}
\label{sec:population}

\rd{We identified the population for our interviews considering participants who can best provide the information needed in order to answer our RQs. 
We selected participants that fulfilled the following criteria:}

\rd{\begin{itemize}
    \item Currently developing software projects. With this criteria, we aim at selecting participants that are still working on software projects. This criteria will exclude persons that had a long experience  but are not working anymore in software development projects (e.g. upper managers)
    \item At least 5 years of experience in developing software projects. We aim at including only practitioners with a minimum level of experience, excluding freshman and newly graduated ones.  
    \item At least 3 years of experience in the domain they are working. We want to consider only practitioners that have a minimum level of experience in the domain they are working, to avoid incongruences due to the lack of knowledge of the domain. 
    \item At least 3 years of experience in deciding which OSS component integrate in the product they develop. 
\end{itemize}
}

\subsubsection{The questionnaire}
\label{ss:questionnaire}

The interviews were based on the same questionnaire adopted in our previous works to elicit the factors considered important for evaluating OSS~\cite{DelBianco2009}\cite{Taibi2015ICSEA1}. 
We organized the questions in the questionnaire adopted for the interviews two sections, according to the types of information we sought to collect:

\begin{itemize}
\item \textit{\textbf{Demographic information}}: In order to define the respondents' profile, we collected demographic background information in relation to OSS\rb{, including the number of years of experience in selecting OSS components to be integrated in the software they develop}. This information considered predominant roles and relative experience. We also collected company information such as application domain, organization’s size via number of employees, and number of employees in the respondents' own team.

\item \textit{\textbf{Factors considered during the adoption of OSS}}: 
Here we asked to list and rank the factors considered during the adoption of OSS software to be integrated in the products they develop, based on their importance, on a 0-to-5 scale, where 0 meant “totally irrelevant” and 5 meant “fundamental”.

\begin{itemize}
\item We first asked to list the factors the respondents consider when adopting an OSS in the software product they develop, and to rank the them on the 0-to-5 scale. This open question is to encourage respondents to identify the important factors which might not be clarified in our Euromicro/SEAA \rs{SLR}~\cite{Lenarduzzi2020SEAA}.

    \item Then, we asked to \rs{rank other possible factors not mentioned in the previous step, on the 0-to-5  scale. Please note that the interviewer listed the remaining factors identified in our Euromicro/SEAA SLR and not mentioned by the participant~\cite{Lenarduzzi2020SEAA}.}
     \rs{The factors identified in the Euromicro/SEAA SLR are reported below}. 
            \begin{itemize}
            \item  Community \& Support
            \item  Documentation
            \item  Economic
            \item  License
            \item  Operational SW Characteristics
            \item  Maturity
            \item  Quality
            \item  Risk
            \item Trustworthiness
        \end{itemize}
    
    \item For each factor ranked higher or equal than 3, we asked to:
    \begin{itemize}
        \item Report 
        \rs{the related} sub-factors \rs{ and their associated metrics with the importance ranking } 
        \rs{on  the  0-to-5  scale}
        \item Report the source they commonly use to evaluate them (e.g. GitHub, Jira, manual inspection, ...)
        \item Report the metrics they adopt to measure the factor
    \end{itemize}

    \item We finally asked if they think the factors they reported enable a reasoned selection of OSS or if they would still need some piece of information to have a complete picture of the assessment.  
\end{itemize}
\end{itemize}

The complete questionnaire adopted in the interviews is reported in the replication package~\cite{ReplicationPackage}.

\subsubsection{Interviews Execution}
\label{StudyExecution}
The interviews were conducted online, using different videoconferencing tools (Zoom, Skype and Microsoft Teams), based on the tool preferred by the interviewed participant. Interviews were carried out from September 2020 to December 2020. 

Because of time constraint, and of the impossibility to conduct face-to-face interviews during public events, interviewees were selected using a convenience sampling approach (also known as Haphazard Sampling or Accidental Sampling)~\cite{battaglia2008convenience}.   
However, we tried to maximize the diversity of the interviewees, inviting an equal number of developers from large and medium companies, and from companies in different domains. The selected participants are experienced developers or project managers, and have been involved in the OSS selection process or the software integration and configuration management process. 
We did not consider any profiles coming from academia, such as researchers or students, nor any inexperienced or junior profiles.  

\subsubsection{Interviews Data Analysis}
\label{Dataanalysis}

\rt{Nominal data on the factor importance is determined by the proportion of responses in the according category. In order to avoid bias, the interviewees are asked to recall the important factors without being provided with options. Thus, the proportion of interviewees who mentioning a factor shall reflect its importance fairly. Ordinal data, such as 5-point Likert scales, was not converted into numerical equivalents to prevent the risk of misleading to subsequent analysis. Apparently when the deviation of the responses is large, such a phenomenon will be overlooked. In this way, we can better identify the potential distribution of the interviewees' responses.}


Open questions (application domain, other factors reported, platforms adopted to extract the information and metrics adopted to evaluate the factors) were analyzed via open and selective coding~\cite{Wuetherick2010}. The answers were interpreted by extracting concrete sets of similar answers and grouping them based on their perceived similarity.
Two authors manually provided a hierarchical set of codes from all the transcribed answers, applying the open coding methodology~\cite{Wuetherick2010}. The authors discussed and resolved coding discrepancies and then applied the axial coding methodology~\cite{Wuetherick2010}.

\subsection{Step 2: Analysis of the APIs of the OSS portals}


We manually  analyzed the  APIs of the portals identified in RQ3, looking for APIs that allowed to assess the information needed to measure the factors reported by the interviewees (RQ1 and RQ2). 
The first two authors independently analyzed all the portals seeking for these pieces of information, and then compared the results obtained. 
In case of discrepancies, all the incongruities were discussed by all the authors, reaching a 100\% consensus. 

Some factors were not directly analyzable. \rb{For example, the responsiveness of an OSS community cannot be directly measured; hence, a proxy metric, i.e., the average time spans between the created time of issues and the first actions, is adopted.} Therefore, the first two authors proposed a list of proxy metrics, considering both the metrics adopted by the interviewees  and metrics available in the literature. Then, all metrics were discussed by all the authors until we reach a consensus. 

However, as expected, not all the metrics can be automatically extracted, and some of them require a manual assessment. An example of a factor that cannot be automatically extracted is the availability of complete and updated architectural documentation.

\subsection{Step 3: Analysis of projects that provide information to assess the factors}




\subsubsection{ Validation of the factors analyzability on the OSS portals }
This step was based on three sub-steps:

\begin{itemize}

    \item \textit{Project selection}. We selected the top 100K GitHub projects, based on the number of stars. \rf{The list of selected projects were determined on 2020-11-10.} The number of projects was limited to the time available. In particular, the different APIs limit the number of queries that can be executed in one hour, and therefore we limited the study to 100k most starred projects to ensure that the data 
    can be extracted in 2 months. \rf{We are aware that some projects might not be code-based projects, and some repositories might only have the purpose to collect resources. However, since it is not possible to automatically exclude non-code projects, we consider them all. }
    \item \textit{Information extraction}. We extracted the selected  information from the APIs. We decided to extract only the information needed to evaluate the factors. Other information are available, but requires to run a higher number of queries, and therefore would have reduced the number of projects that we can extract. \rf{The extraction process started on 2020-11-16 with data collected gradually till 2020-12-29.} As an example, it would be possible to extract all the details on project issues (issue title, author, date, comments, ...), but this would have required to run a number of additional queries, without providing any  information considered valuable by our interviewees. 
    \item \textit{Analysis of the information available}. In this step we analyzed which information is actually available for each project. 
    As an example, not all the projects might use different issue trackers instead of using the one provided by GitHub, or some projects might not be listed by the NIST NVD database, or more, some project might not have questions and answers available in StackOverflow or Reddit. 
    
\end{itemize}

\subsection{Replication}

In order to ease the replication of this work we provide the complete replication package including the questionnaire adopted for the interviews, the results obtained in the interviews, the data crawling script and the results of the data analysis~\cite{ReplicationPackage}. 

\label{sec:Method}

\section{Results}
Here we first provide information about the sample of respondents, which can be used to better interpret the results and then, we show the collected results with a concise analysis of the responses obtained, with insights gained by statistical analysis.

\begin{figure}[t]
\centering
\subcaptionbox{}
{\begin{tabular}[c]{|p{2.5cm}|l|} 
         \hline
        Company Size (\# Employees)       & \# Interviewees \\ 
        \hline
        0-50               & 4             \\
        50-250             & 3             \\
        250-1000           & 7             \\
        \textgreater{}1000 & 9            \\ 
        \hline
      \end{tabular} }
\subcaptionbox{Distribution of company size}
{\includegraphics[width=0.4\linewidth,valign=b]{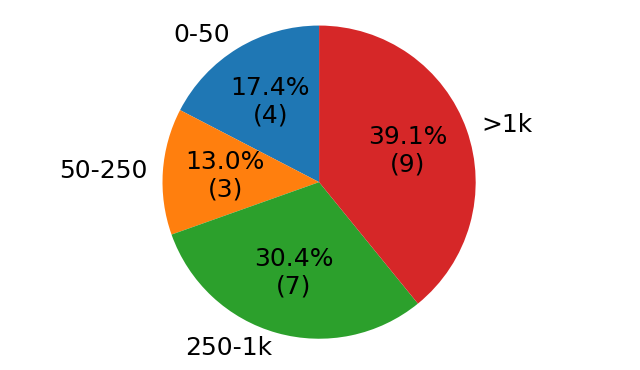}}
\caption{Distribution of company sizes of our interviewees}
\label{img:sizes}
\end{figure}

\begin{figure}[t]
\centering
\subcaptionbox{}
{\begin{tabular}[c]{|l|l|} \hline
        Roles              & \# Interviewees  \\ \hline
        SW developer       & 9 \\
        Product Manager    & 8 \\
        Software Architect & 6 \\ \hline
        \end{tabular}}
\subcaptionbox{The roles of our interviewees}
{\includegraphics[width=0.4\linewidth,valign=b]{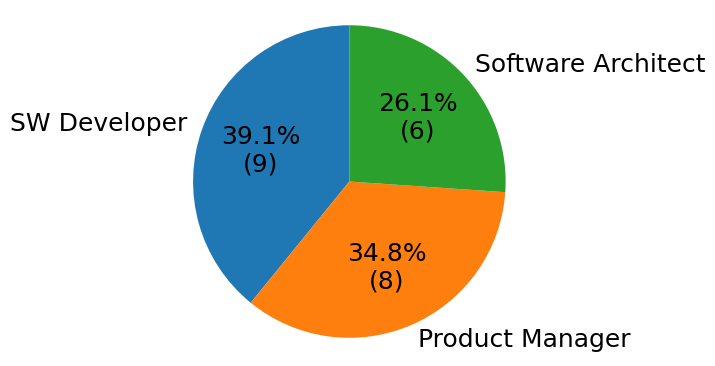}}
\caption{Figure caption}
\label{img:roles}
\end{figure}

\begin{figure}[ht!]
\centering\includegraphics[width=0.8\linewidth]{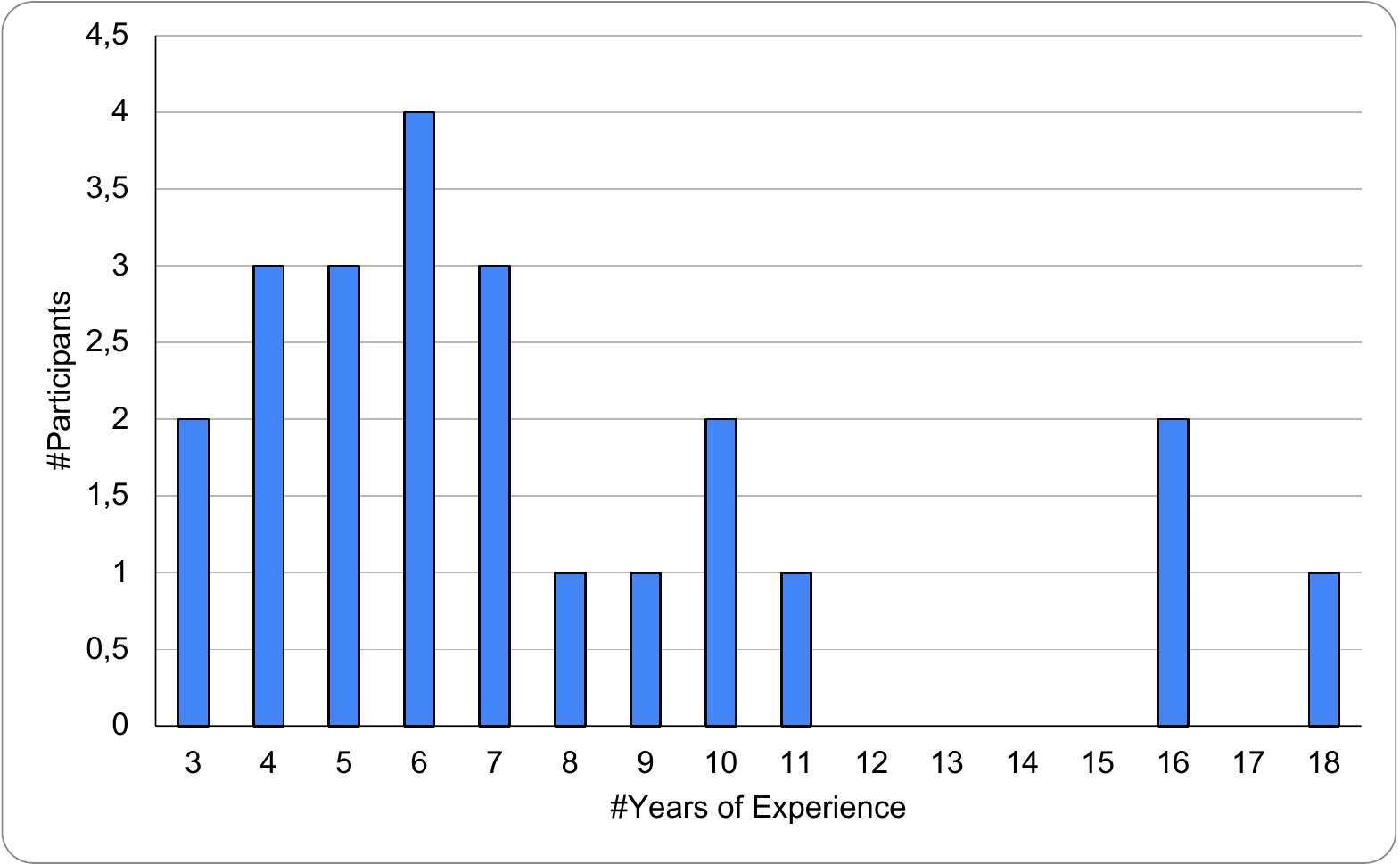}
\caption{Number of years of experience of the participants in selecting OSS components to be integrated in the products they develop}
\label{fig:experience}
\end{figure}

We collected 23 interviews from experienced practitioners. 
Figure~\ref{img:sizes} contains the distribution of company sizes where our interviewees belong,  Figure~\ref{img:roles} shows the percentage for organizational roles identified in the questionnaire \rd{ while Figure \ref{fig:experience} shows the distribution of the experience of our interviewees in selecting OSS components to be integrated in the projects they develop}.


\subsection{RQ1: Factors considered by practitioners when selecting OSS}

Our interviewees consider 8 main factors and 46 sub-factors when they select OSS, reporting an average of 2.35 factors per interviewee, a minimum of 1 and a maximum of 21 factors.

The factor that is mentioned more frequently from the interviewees is \textit{License} which has received a median importance of 4 out of 5. Surprisingly, this is not the value which has received the highest median value of importance as \textit{Community Support and Adoption, Performances} and \textit{Perceived Risk} received a median value of importance of 4.5 out of 5. 
It is interesting to note that no participant mentioned \textit{economic} \rs{and its} related \rs{sub-}factors such as license costs, or cost for training. \rs{So this factor is not reported in our results.} 

In Table~\ref{tab:interviewees-portals-factors}, we report the list of factors and sub-factors together with the number of participants who mentioned them (column RQ1- \#) and the median of the importance reported by the interviewees (column RQ1 - Median).

\begin{table}
\caption{High-level Factors considered during the adoption of OSS}
\label{tab:interviewees-portals-factors}
\centering
\scriptsize
\adjustbox{center}{
\begin{tabular}{|ll|p{0.2cm}|c|c|} \hline

\multicolumn{4}{|l|}{	\textbf{RQ1}	} &	 \textbf{RQ2}	
\\ \hline
\multicolumn{2}{|l|}{	Factor	}		&			\#	&	Median	&			\#Metrics	
\\	 \hline
\multicolumn{2}{|l|}{	Community Support and Adoption	}		&			10	&	4.5	&				
\\	  
		&	Popularity	&			9	&	3	&			4	
		\\

		&	Community reputation	&			11	&	3	&			3	\\

		&	Community size	&			13	&	3	&			5	\\

		&	Communication	&			6	&	3.5	&			5	\\

		&	Involvement	&			9	&	3	&			1	\\

		&	Sustainability	&			11	&	3	&			1	\\

		&	Product Team	&			5	&	3	&			2	\\

		&	Responsiveness	&			1	&	5	&			1	\\

																			  \hline
\multicolumn{2}{|l|}{	Documentation	}		&			14	&	4	&			\\
		&	Usage documentation	&			4	&	4	&			5	\\

		&	Software requirements	&			11	&	3	&			1	\\
																			
		&	Hardware requirements	&			8	&	3.5	&			1	\\
																			
		&	Software Quality Documentation	&			5	&	3	&			3	\\

																			  \hline
\multicolumn{2}{|l|}{	License	}		&			21	&	4	&			7	\\
\hline

\multicolumn{2}{|l|}{	Operational SW Characteristics	}		&			6	&	4	&		\\
		&	Trialability	&			5	&	3	&			2	\\

		&	Independence from other SW	&			11	&	3	&			4	\\

		&	Development language	&			5	&	4	&			3	\\

		&	Portability	&			1	&	4	&			1	\\
																			
		&	Standard compliance	&			5	&	4	&			0	\\

		&	Testability	&			6	&	3.5	&			0	\\

																			  \hline
\multicolumn{2}{|l|}{	Maturity	}		&			6	&	3.5	&			11	\\
\hline

\multicolumn{2}{|l|}{	Quality	}		&			6	&	3.5	&				\\
																			
		&	Reliability	&			3	&	4	&			6	\\

		&	Performances	&			4	&	4.5	&			1	\\

		&	Security	&			15	&	4	&			6	\\

		&	Modularity	&			3	&	3	&			1	\\

		&	Portability	&			3	&	4	&			2	\\

		&	Flexibility/Exploitability	&			3	&	3	&			3	\\

		&	Code Quality	&			13	&	4	&			6	\\

		&	Coding conventions	&			9	&	3	&			0	\\

		&	Maintainability	&			3	&	4	&			0	\\
																			
		&	Testability	&			2	&	4	&			0	\\
		&	Existence of benchmark/test	&			4	&	3.5	&			4	\\

		&	Changeability	&			2	&	3.5	&			0	\\

		&	Update/Upgrade/Add-ons/Plugin	&			3	&	4	&			1	\\

		&	Architectural quality	&			5	&	3	&			0	\\

																			  \hline
\multicolumn{2}{|l|}{	Risk (Perceived risks)	}		&			7	&	4.5	&				\\
		&	Perceived lack of confidentiality	&			5	&	1	&			0	\\
		&	Perceived lack of integrity	&			5	&	3	&			0	\\
		&	Perceived high availability	&			5	&	4	&			3	\\

		&	Perceived high structural assurance	&			5	&	2	&			0	\\
																			
		&	Strategic risks	&			5	&	3	&			0	\\
		&	Operational risks	&			5	&	1	&			1	\\
																			
		&	Financial risks	&			5	&	2	&			0	\\
		&	Hazard risks	&			5	&	4	&			5	\\

																			  \hline
\multicolumn{2}{|l|}{	Trustworthiness	}		&			6	&	4	&				\\
		&	Component	&			4	&	3.5	&			3	\\

		&	Architecture	&			4	&	3	&			2	\\

		&	System	&			4	&	3.5	&			3	\\

		&	OSS provider reputation	&			4	&	3.5	&			0	\\
																			
		&	Collaboration with other product	&			4	&	2.5	&			3	\\

		&	Assessment results from 3rd parties	&			2	&	3.75	&			0	\\

																			 \hline
\end{tabular}
}
\end{table}

\subsection{RQ2: Metrics used by practitioners to evaluate factors during OSS selection}

When we asked practitioners to report the metrics they use to evaluate the factors they mentioned in RQ1, and to rank their usefulness, practitioners mentioned 110 different metrics.

The complete list of metrics reported for each factor is reported in Appendix A. 

In Table~\ref{tab:interviewees-portals-factors} (Column RQ2 - \#Metrics) we report the count of metrics considered as useful by practitioners (likert scale $\geq$ 3, where 0 means ''This metric is useless to evaluate this factor'' and 5 means that the metric is extremely useful). 
 
Surprisingly, the factor where practitioners provided the highest number of metrics to assess it, is \textit{Maturity}, which has been mentioned only 6 times compared to the \textit{License}, mentioned 21 times,  where practitioners provided 7 metrics instead. This indicates a wide variety of interpretations on  \textit{Maturity}, and practitioners use different metrics to evaluate this factor. The careful reader can also observe that for some factors considered as relevant in RQ1 such as \textit{Perceived risks}, no metrics have been mentioned. This result proves that in some cases, some of the most important factors in an OSS can not be objectively measured and the interviewees do not know how to retrieve such information appropriately.

\subsection{RQ3: Sources of information and portals used to assess OSS}
Our interviewees mentioned 9 different source of information and portals they commonly consider when they select OSS.

\begin{figure}[ht!]
\centering\includegraphics[trim={5cm 16.5cm 5cm 4.5cm},clip,width=\linewidth]{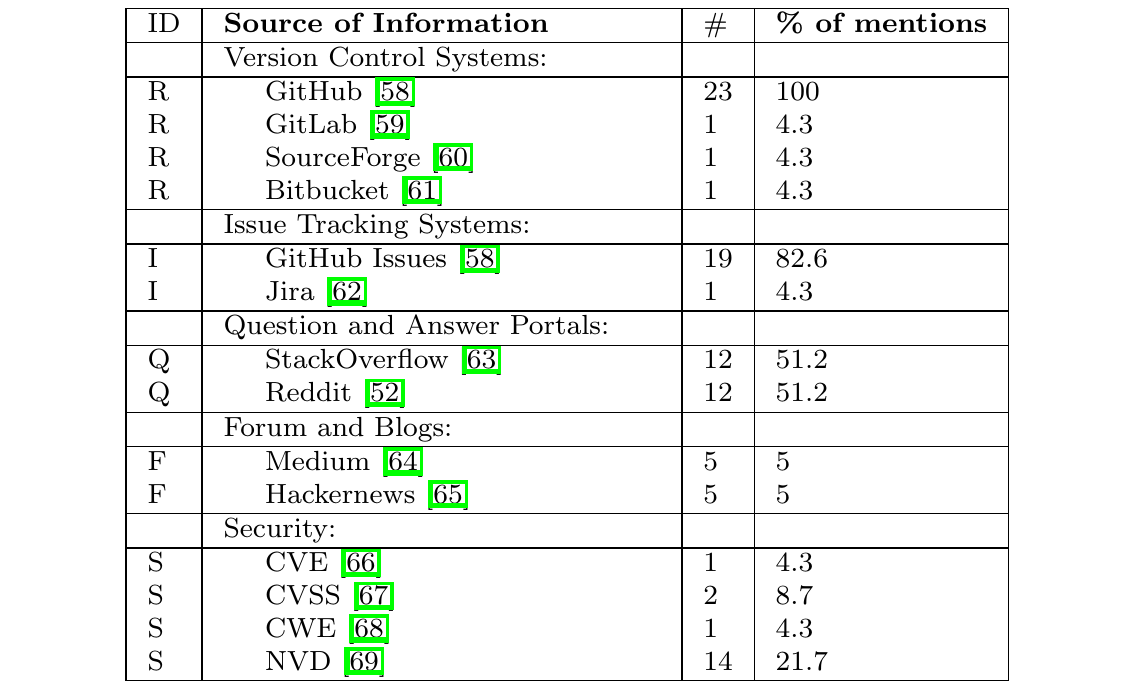}
\captionof{table}{The source of information reported by the interviewees (RQ3)}
\label{tab:SoI}
\end{figure}

In Table~\ref{tab:SoI} we list the sources of information adopted by the practitioners to evaluate OSS, together with the number of participants who mentioned it (columns \# and \% of mentions). To increase the readability, we grouped the source of information in five main categories: \textit{version control systems}, \textit{issue tracking systems}, \textit{Question and Answer portals} (Q\&A), \textit{forum and blogs} and \textit{security} related platforms. 

The 5 most reported sources of information are GitHub (and GitHub Issue tracker), StackOverflow, Reddit, and NIST Security Vulnerability (NVD) with respectively: 23, 19, 12, 12 and 14 mentions. All of these are mentioned from more than 20\% of the interviewees and are therefore those which prove to be the most useful when retrieving information related to OSS. Other platforms such as Bitbucket or Jira were rarely mentioned. \rt{The results presented in Table~\ref{tab:SoI} could also be useful to other OSS stakeholders such as software administrators and software operators.}


\subsection{RQ4: Factors that can be extracted automatically from the portals}

We focused our attention to the four most used sources of information reported in RQ3: GitHub~\cite{GitHub}, StackOverflow~\cite{StackOverflow}, Reddit~\cite{reddit} and the NIST National Vulnerability Database~\cite{NVD}. 
For such purpose, we first identified the  APIs that can be adopted to extract the information needed to measure the factors, and then we extracted the data from the 100k with more stars in GitHub.  

The extraction of the information for 100K projects took a total of 5 days for GitHub, 53 days for StackOverflow, 4 days for Reddit and 2 Days for the NIST National Vulnerability Database. 
The long processing time is due to the limit of queries that can be performed on the APIs for different IP addresses. 

\begin{figure}
\centering
\begin{subfigure}{.48\textwidth}
  \centering
  \includegraphics[width=\linewidth]{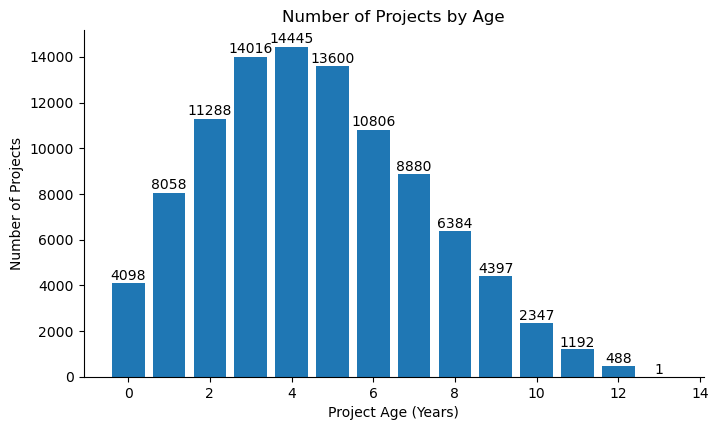}
  \caption{OSS Project Stats by Age (Years)}
  \label{fig:age}
\end{subfigure}%
\begin{subfigure}{.48\textwidth}
  \centering
  \includegraphics[width=\linewidth]{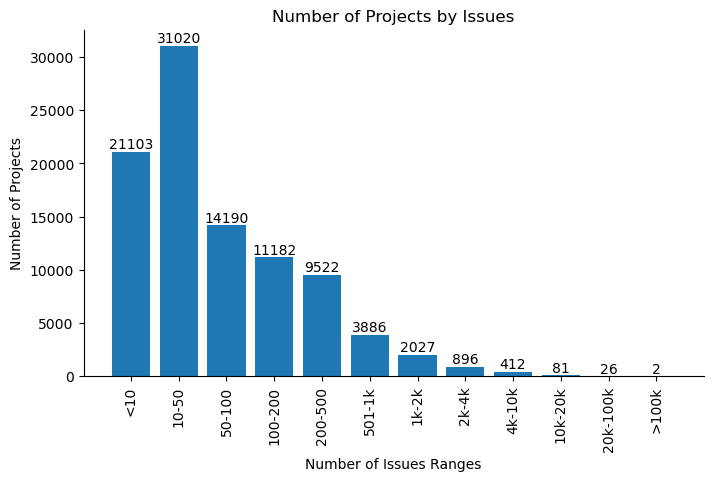}
  \caption{OSS Project Stats by Issues}
  \label{fig:issue}
\end{subfigure}
\begin{subfigure}{.48\textwidth}
  \centering
  \includegraphics[width=\linewidth]{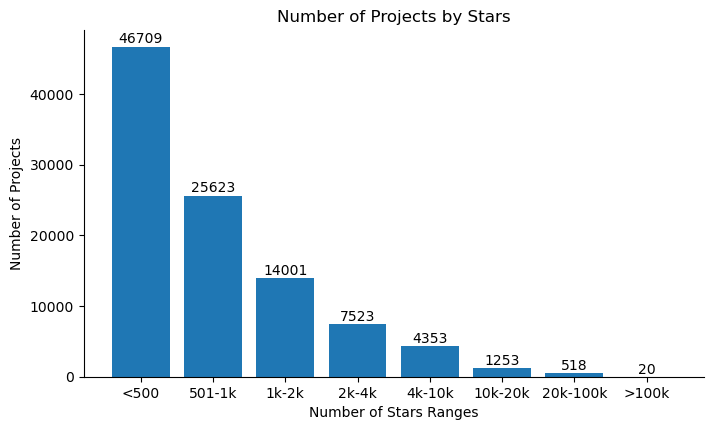}
  \caption{OSS Project Stats by Stars}
  \label{fig:star}
\end{subfigure}
\begin{subfigure}{.48\textwidth}
  \centering
  \includegraphics[width=\linewidth]{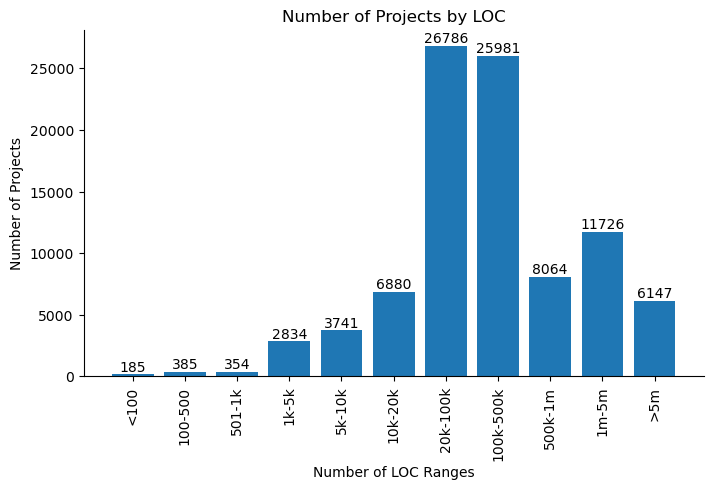}
  \caption{OSS Project Stats by LOC}
  \label{fig:loc}
\end{subfigure}
\begin{subfigure}{.48\textwidth}
  \centering
  \includegraphics[width=\linewidth]{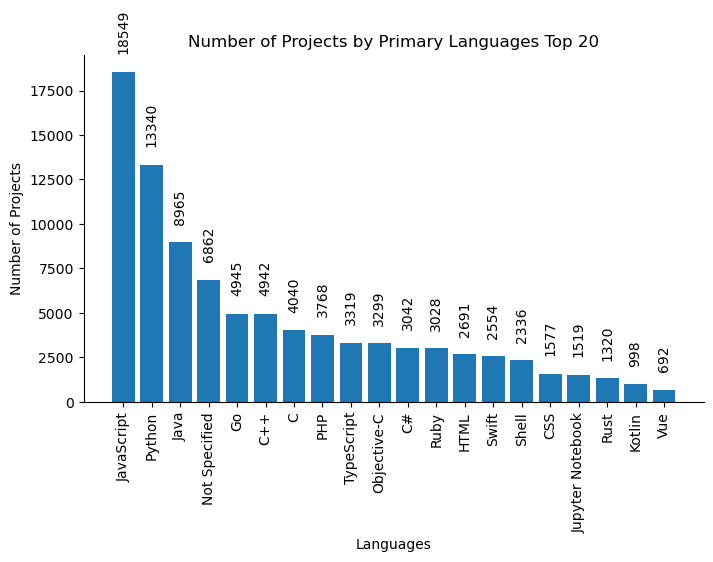}
  \caption{OSS Project Stats by Top 20 Primary Languages}
  \label{fig:pl}
\end{subfigure}
\begin{subfigure}{.48\textwidth}
  \centering
  \includegraphics[width=\linewidth]{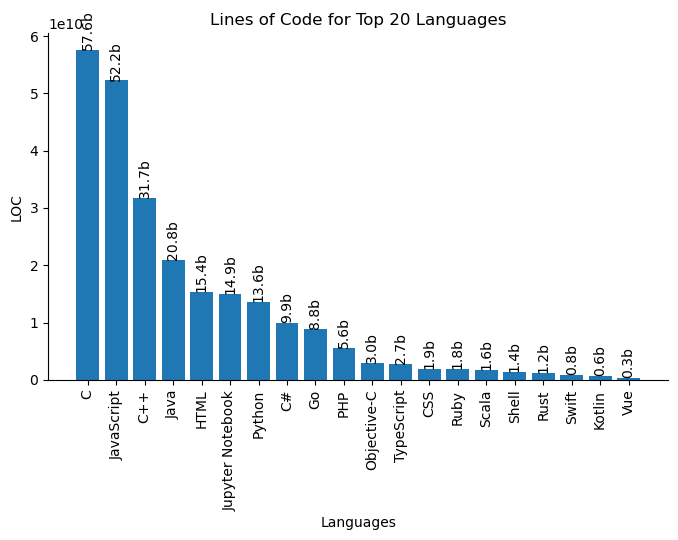}
  \caption{Stats for Top 20 Language by LOC}
  \label{fig:langloc}
\end{subfigure}
\begin{subfigure}{.48\textwidth}
  \centering
  \includegraphics[width=\linewidth]{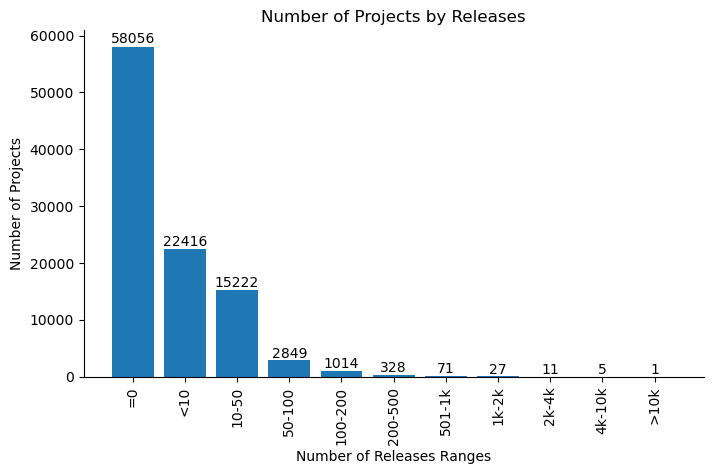}
  \caption{OSS Project Stats by Releases}
  \label{fig:release}
\end{subfigure}
\begin{subfigure}{.48\textwidth}
  \centering
  \includegraphics[width=\linewidth]{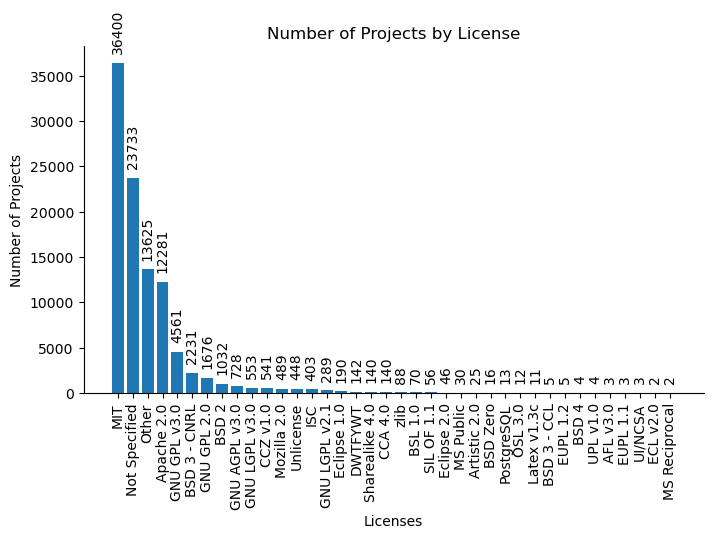}
  \caption{OSS Project Stats by Licenses}
  \label{fig:license}
\end{subfigure}
\caption{General Stats for Top 100k Github Projects}
\label{fig:gstats}
\end{figure}

Considering the projects extracted (Figure \ref{fig:age}), more than half of these projects (53.3\%) have been active for 2 - 6 years. Around 
12.1\% of these projects are active for one year or less when only 3.9\% of them are active for more than 10 years. Majority of these projects (87.0\%) have less than 500 issues during their life cycle when around 3.4\% of them have more than 1k issues (Figure \ref{fig:issue}). Furthermore, there are 1791 projects being very popular having more than 10k stars when 25.9\% projects having stars ranging from 1k to 10k (Figure \ref{fig:star}) with 46.7\% having less than 500 stars. On the other hand, regarding project size, more than half of them (52.8\%) have lines of code (LOC) ranging from 20k to 500k. 6.1\% projects contain more than 5m LOC when only 0.9\% of them have less than 1k LOC (Figure \ref{fig:loc}). Regarding developing languages, Javascript is the most popular being the primary language of OSS projects (17k projects) with Python and Java at 2nd and 3rd (Figure \ref{fig:pl}). They are the primary languages for 40.9\% of the projects. However, regarding LOC by languages, C language (57.6b) and Javascript (57.2b) rank at the top. Both have almost doubled the amount of C++ language (31.7b) which ranks the 3rd in terms of total LOC (Figure \ref{fig:langloc}). Regarding release numbers, 58.1\% projects do not have any specific release recorded. 37.6\% have less than 50 releases when only the rest 4.3\% have more than 50 releases. 
All the previously mentioned data is always available from GitHub, and queries to the GitHub APIs will always return a valid information. 

Regarding the adopted open source licenses, 23.7\% projects did not specify the licenses they adopt. As for the other projects, MIT, Apache 2.0 and GNU GPL v3.0 are the most popular licenses with 53.2\% projects adopting one of them (Figure \ref{fig:license}). Therein, 13.6\% projects adopted non-mainstream license (identified as 'Other').

\begin{figure}
\centering
\begin{subfigure}{.48\textwidth}
  \centering
  \includegraphics[width=\linewidth]{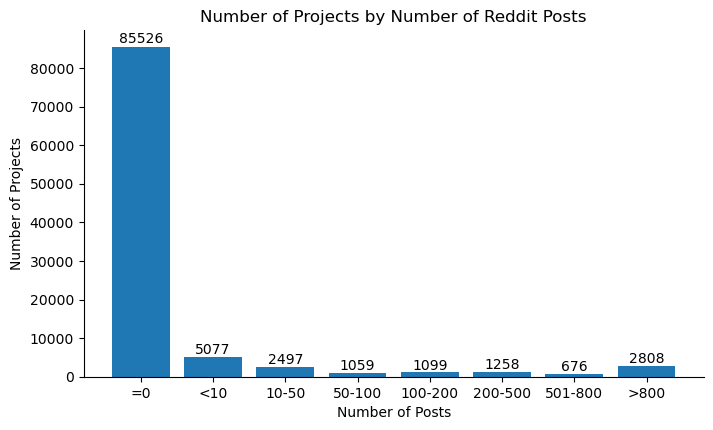}
  \caption{OSS Project Stats by Reddit Post Number}
  \label{fig:reddit}
\end{subfigure}%
\begin{subfigure}{.48\textwidth}
  \centering
  \includegraphics[width=\linewidth]{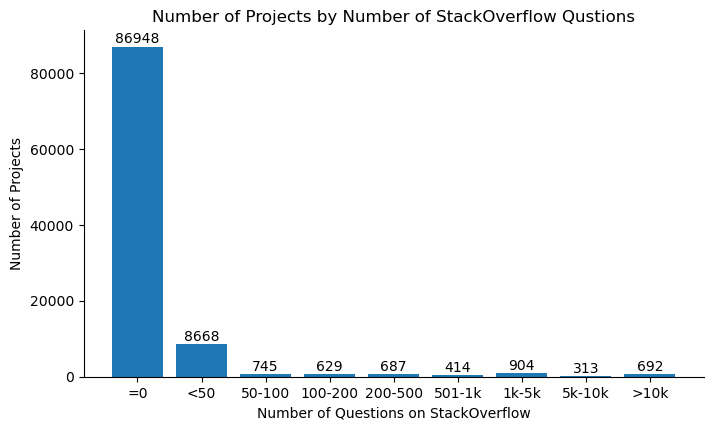}
  \caption{OSS Project Stats by StackOverflow Questions}
  \label{fig:stack}
\end{subfigure}
\caption{Stats for Top 100k Github Projects in Reddit and StackOverflow}
\label{fig:rsstats}
\end{figure}

We also validate the APIs of Reddit and StackOverflow by finding the amount of discussion threads on each of the 100k OSS projects. As shown in Figure~\ref{fig:rsstats}, 14.5\% projects are generally discussed in Reddit with  only 5.8\% projects having more than 100 posts (Figure \ref{fig:reddit}). On the other hand, 13.0\% projects are discussed (raised technical questions) in StackOverflow (Figure \ref{fig:stack}). Therein, 3.6\% projects have more than 100 questions raised on. In general, such results show that it is hard to find sufficient generic or technical discussion regarding specific OS projects from Reddit and StackOverflow.

Based on the interview results, especially the obtained factors that are considered important by the practitioners (shown in Table~\ref{tab:interviewees-portals-factors}), we further validate whether such factors can be analyzed via the automatically obtained data from the previously mentioned portals with the results shown in Table~\ref{tab:factor_available}.

\begin{table}[]
\caption{Factor Information Availability Stats for Top 100K Projects}
\label{tab:factor_available}
\adjustbox{max width=\textwidth}{
\centering
\scriptsize
\begin{tabular}{|l|l|l|l|} \hline
\textbf{Factor \rs{(\#part-auto/\#full-auto/\#metrics)}}&\textbf{Portal}&\textbf{\#}&\textbf{\%}\\ \hline
Community Support and Adoption \rs{(5/16/41)} &&&\\

\hspace{2em} Popularity &R&&		\\
\hspace{4em} Number of Watch	&	R	&	100000&	100.00\%		\\
\hspace{4em} Number of Stars	&	R	&	100000&	100.00\%		\\
\hspace{4em} Number of Forks	&	R		&100000	&100.00\%		\\
\hspace{4em} Number of Downloads	&	R	&	42260&	42.26\%		\\

\hspace{2em} Community reputation &I+*&&\\
\hspace{4em} Fast response to issues \rb{($\triangle$)} &I+*	&	100000&	100.00\%	\\

\hspace{2em} Community size	&	R+*	&	&		\\
\hspace{4em}		Number of Contributors	&	R*		&100000	&100.00\%		\\
\hspace{4em}		Number of Subscribers	&	R		&100000	&100.00\%		\\
\hspace{4em}		Community age	&	R+		&100000	&100.00\%		\\
\hspace{4em}		Number of Involved developers per company	&	R+*		&100000	&100.00\%		\\
\hspace{4em}		Number of Independent developers	&	R+*	&	100000&	100.00\%\\
\hspace{2em} Community support &	R+*, Q+*		&	&	\\
\hspace{4em}		Activeness	\rb{($\triangle$)}&	R+*		&100000&	100.00\%		\\
\hspace{4em}		Responsiveness	\rb{($\triangle$)}&	R+*, Q+*		&14474&	14.47\%		\\
\hspace{2em} Communication	&		I+*, Q+*	&&	\\
\hspace{4em}	Availability of questions/answers	&	I, Q	&	100000&	100.00\%		\\
\hspace{4em} Availability of forum	&	Q	&	14474	&14.47\%		\\
\hspace{4em}		Responsiveness of postings	&	I+*, Q+*	&	14474&	14.47\%		\\
\hspace{4em}		Quality of postings	\rb{($\triangle$)}&	Q+*	&	14474&	14.47\%		\\
\hspace{2em}Sustainability		&	I+*		&		&			\\
\hspace{4em}		Existence of maintainer	&	I+*		&	100000	&	100.00\%		\\

\hspace{2em} Product Team		&	R+*		&		&			\\
\hspace{4em}		Developer quality \rb{($\triangle$)}&	R+*		&	100000	&	100.00\%		\\
\hspace{4em}		Developer Productivity \rb{($\triangle$)}&	R+*		&	100000	&	100.00\%		\\
\hspace{2em} Responsiveness	&		I+*		&		&			\\
											
\hspace{4em}		Avg. bug fixing time	&	I+*		&		100000	&	100.00\%			\\
\hspace{4em}		Avg time to implement new issues	&	I+*		&		100000	&	100.00\%			\\

Documentation \rs{(3/0/12)}&&&\\
\hspace{2em} Usage documentation	&		R*, F+		&		&			\\
\hspace{4em}		Availability of updated documentation	\rb{($\triangle$)}&	R*		&	42260	&	42.26\%		\\
\hspace{4em}		Availability of documentation/books/online docs	\rb{($\triangle$)}&	R		&	39499	&	39.50\%		\\
\hspace{4em}		Availability of Tutorial or Examples	\rb{($\triangle$)}&	R		&	39499	&	39.50\%		\\
License	\rs{(1/0/9)}&	R		&		&			\\
\hspace{2em} License type	&	R		&	76576	&	76.58\%		\\
Operational SW Characteristics	\rs{(1/0/9)}&	R		&		&			\\
\hspace{2em} Development language		&	R		&	93148	&	93.15\%		\\
Maturity \rs{(4/3/11)}&	R+*		&		&			\\

\hspace{4em} Number of forks	&	R		&	100000	&	100.00\%		\\
\hspace{4em} Release frequency	&	R+*		&	42260	&	42.26\%		\\
\hspace{4em} Number of releases	&	R		&	42260	&	42.26\%		\\
									
\hspace{4em} Age (in Years)	&	R+		&	100000	&	100.00\%		\\
\hspace{4em} Number of commits	&	R		&	100000	&	100.00\%		\\
									
\hspace{4em} Development versions	&	R*		&	42260	&	42.26\%		\\
\hspace{4em} New feature integration	&	R+*		&	42260	&	42.26\%		\\

Quality	\rs{(5/3/47)}&		&		&			\\
\hspace{2em} Reliability &	I+*		&		& \\
\hspace{4em} Average bug age	&	I+*		&	100000	&	100.00\%		\\
\hspace{4em} Avg. bug fixing time	&	I+*		&	100000	&	100.00\%		\\

\hspace{2em}Security		&	R+*, Q+*, S+*		&		&			\\
\hspace{4em}	Number security vulnerabilities	&	S		&12838 &	12.84\%			\\
\hspace{4em}	Number of Vulnerabilities reported on the NVD portal	&	S		&	12838	&	12.84\%		\\

\hspace{4em}	Vulnerability Resolving time \rb{($\triangle$)}&	R+*, S+		&	12838	&	12.84\%		\\
\hspace{4em}	Community concern towards security	\rb{($\triangle$)}&	R+*, Q+*, S+*		&	12838	&	12.84\%		\\
\hspace{4em}	Vulnerability impact \rb{($\triangle$)}&	S		&	12838	&	12.84\%		\\
\hspace{2em} Code Quality &R&&\\
\hspace{4em} Code size	&	R		&	100000	&	100.00\%		\\
Trustworthiness \rs{(1/0/22)} &R+*&&\\
\hspace{4em} Consistent release updates pace \rb{($\triangle$)}&	R+*		&	42260	&	42.26\%		\\
\hline
\end{tabular}
}
\begin{tablenotes}
    \centering
    \tiny
    \item\textit{Repository(R), Issue Tracker(I), Questions and Answer portals(Q), Forum \& Blogs(F), Security(S)}
    \item\textit{Calculation Required(+), Multi-queries Required(*), \rb{Proxy Metrics($\triangle$)}}
\end{tablenotes}
\end{table}

According to the investigation on automatic OSS data extraction on the Top 100k OSS projects on Github via the APIs of the five public portals, we find that the majority of the metrics towards the \textit{Community Support and Adoption} factor can be automatically done via data extractions from such APIs. To be noted, regarding \textit{Communication}, the question and answer sources (i.e., StackOverflow) contain information on limited number of OSS projects, which limits the availability towards the evaluation on such category. In addition, we also find, despite the availability of automatic data extraction and measuring via APIs, many of the metrics require further calculation and learning, as well as multiple queries to obtain. For example, in order to measure the \textit{Number of Independent Developers}, we must get the list of contributors of a particular project via multiple queries first (max items per page for Github API is 100), and check the ``Independentness'' of each contributor via further investigating his/her organization status. Thus, such process shall be, to some extent, time-consuming, when the limit rate of the API usage shall be also taken into account. 

Another category can be automatically measured is \textit{License}, as shown in the data, 76.6\% of the projects contain specific License information. On the other hand, for the other categories, automatic data extraction and full-grained evaluation is hindered by the limited availability of the according data, as only very limited percentage of the metrics can be automatically done via data extraction (shown in Table \ref{tab:interviewees-portals-factors}). And amongst these categories, the evaluation of \textit{Maturity} category depends on the availability of the release data from repository dataset, when for the obtained 100k projects only 42.2\% provide such information. Measuring the availability of \textit{Documentation} shall also depend on the information extracted from project description and homepage, while only around 40\% of projects provide those. 

As for the security vulnerability evaluation, the NVD Dataset provide information for 12838 projects (12.8\%). However, it is not clear if the projects not reported in NVD do not contain security vulnerabilities at all, or simply are not indexed by the NVD dataset. However, it is important to note that the NVD performs analysis on CVEs published to the CVE Dictionary. Every CVE has a CVE ID which is used by cybersecurity product/service vendors and researchers for identifying vulnerabilities and for cross-linking with other repositories that also use the CVE ID. However, it is possible that some security vulnerabilities are not publicly reported with assigned CVE IDs.

\rs{As shown in Table \ref{tab:factor_available}, amongst the 170 metrics identified, 40 of them are potentially available to be extracted automatically. In addition, \textit{License type} and \textit{Development Language}, though seen as sub-factors, can be automatically measured as well. Therein, the number of automatic measurable metrics for all 100k projects (\#full-auto) and that for part of them (\#part-auto) are also shown. Only 22 metrics out of 170 can be obtained automatically for all 100k projects when the others are only available for part of the projects. In addition, 22 metrics require multi-queries to complete when 21 require further calculation and/or learning to determine. }

\label{sec:Results}

\section{Discussion}
In this Section, we discuss the results of our RQs and we present the threats to validity of our work.

The factors  and the metrics adopted to evaluate and select OSS (RQ1 - RQ2) evolved over time. While in 2015~\cite{Taibi2015ICSEA1}\cite{Lenarduzzi2020SEAA} factors such as \textit{Customization easiness} and \textit{Ethic} were the most important, nowadays we cannot state the same. Already in 2020~\cite{Lenarduzzi2020SEAA} such factors have been incorporated inside other more valuable factors such as \textit{Quality}, while today are not mentioned anymore. On the other side \textit{License} and \textit{Documentation}, which are the most mentioned factors nowadays were side factors in 2020. As a matter of fact the latter was a sub-section of \textit{Development Process}, while both were not even considered in 2015. Moreover, ethical principles, that were very relevant in 2015, are not even mentioned in 2020. 

Nowadays the trend is to search for OSSs which are ready to be integrated as is. In order to incorporate OSSs without falling into lawsuit particular attention needs to be put into the \textit{License type}, while to guarantee the correct functioning the focus needs to be put in the documentation. A clear example is the necessity of a clear definition of the \textit{system requirements} when incorporating libraries.

Another factor which gained a lot of importance over time is \textit{Security}. While in 2015 was a factor with medium relevance, nowadays it is a keypoint for measuring quality of an OSS. The growing number of portals dedicated to ensure absence of vulnerabilities proves that people are concerned of the use of OSSs when embedded in their system. In particular they strongly rely on such portals to check the history of the OSSs to incorporate and in some cases also to ensure that proper reports are delivered when a new vulnerability is discovered. 

\rt{Also, the importance of a factor is not necessarily proportional to the number of metrics. When specifying the importance of the assessment factors, we may see that some measurable factors are perhaps just eliminated, while some important factors can not be automatically evaluated using the extracted information and require a manual assessment. }
\rf{Moreover, it is important to note that the lack of concrete metrics for some factors, such as community reputation or  community sustainability might be because these factors are too abstract.  Some researchers already addressed some of these aspects. As an example, Cai and Zhu \cite{cai2016reputation} and Hu et al. \cite{hu2012reputation} already proposed some metrics to evaluate the community reputation while Gamalielsson and Lundell \cite{gamalielsson2014sustainability} also identified approaches for contributing to the community sustainability. However, these models are not yet diffused in industry, and this might be the reason why our interviewees were not aware of them.}






The source of information adopted to evaluate OSS (RQ3) did not change completely from the previous years. Users are still adopting project repositories and issue trackers as main source of information. Moreover, an effect of the newly introduced factor security, is that now the selection also require security related information, that are commonly fetched from security databases such as the NIST NVD~\cite{NVD}, CVE~\cite{CVE}, and CWE~\cite{CWE}.  In addition, many vendors like Synopsys\cite{synopsys} and WhiteSource \cite{WhiteSource} offer software composition analysis solutions that facilitate licence risk management, vulnerability identification and management, risk reporting, etc.

Unexpectedly, even nowadays, not all the portals can provide complete information for evaluating the information needed by the practitioners (RQ4). The analysis of the 100K most starred projects in GitHub showed that only the information coming from the project repositories (e.g GitHub) are always available, except for the license information (76.6\%). Considering other factors such as the communication, the situation does not improve, and only in 14.5\% of projects we were able to automatically extract the relevant information from their APIs. \rf{For example, it is noticeable that the APIs of StackOverflow enable the extraction of other information, e.g., the textual content of questions and answers, the users' reputation, and so on. However, it requires further learning and calculation to elicit additional information from such textual content. The possibilities towards such directions shall be studied further in our future studies.} \rt{Furthermore, some of the Github views, though providing valuable information but being inaccessible directly from APIs, (e.g., the GitHub insight view) are not covered herein.} Due to the diversity of application domains, organizational needs, and constraints, practitioners in different organizations may explain the factors from their own perspective and may adopt different metrics in the OSS evaluation. A good example are the metrics associated with the factor of \textit{Maturity}. Metrics such as the number of releases and the system growth in the roadmap were commonly concerned in the evaluation of software maturity; besides, some practitioners also took the number of commits and the number of forks as important metric in the evaluation. The total number of commits itself might not be enough when evaluating software maturity. The prevalence of commits over time and the types of commits could be additional and useful information in the evaluation. Therefore, how the metrics help with the evaluation of the factors could have been clarified.

The results of this work show a discrepancy between the information required by the practitioners to evaluate OSS and the information actually available on the portals, confirming that the collection of the factors required to evaluate an OSS project is very time-consuming, mainly because most of the information required is not commonly available on the OSS project~\cite{Yasutaka2018}\cite{Lenarduzzi2020SEAA}\cite{DelBianco2010QUaliSPO}. 

The automation of the data extraction, using portal APIs might help practitioners reducing the collection time and the subjectivity. 
The result of this work could be highly beneficial for OSS producers, since they could check if they are providing all the information commonly required by who is evaluating their products, and maximize the likelihood of being selected. 
The result can also be useful to potential OSS adopters, who will speed-up the collection of the information needed for the evaluation of the product. 

Even in case OSS producers do not enhance their portals by providing the information required by the practitioners to assess OSS, the results of this work could be useful for practitioners that need to evaluate an OSS product.
The list of factors can be effectively used as checklist to verify if all the potentially important characteristics of OSS have been duly evaluated.
For instance, a practitioner could have forgotten to evaluate the trend of the community activity and he/she could adopt an OSS product that has a ``dissolving'' community: this could create problems in the future because of the lack of maintenance and updates. The usage of checklist would allow practitioners to double check if they considered all factors, thus reducing the potential unexpected issues that could come up after the adoption. 

\subsection{Future Research Directions.} As a result of our findings, we propose the following directions for future research in this area.

Focus on the definition of a common tool to automatically extract information needed for the evaluation of OSS, investigating proxy metrics in case direct metrics are not available. 

Definition of refined and customizable models (which may be obtained by merging multiple available approaches) and favor its adoption through rigorous and extensive validation in industrial settings. This could increase the validity of the model and thus its dissemination in industry, where OSS is still not widely adopted. Several models already exist but, according to the results of our previous literature review~\cite{Lenarduzzi2020SEAA}, they have not been strongly validated and, as a consequence, adoption has been limited.

Try to target the models at quality factors that are of real interest for stakeholders. Most of the available models focus on the overall quality of the product, but few of them are able to adequately assess each single factor that composes the overall quality of the OSS product. This can complicate the assessment of OSS products by stakeholders, who are interested in specific quality factors: e.g., developers are likely more interested in reliability or testability aspects~\cite{Morasca2009}~\cite{Morasca2010}~\cite{Morasca2011}, while business people may be more interested in cost or maintenance factors, etc.. 
 
In the studies, we identified 170 metrics to measure the factors for OSS evaluation and selection, based on which we shall conduct an in-depth analysis to gain a better understanding of the rationale for the metrics. The rationale explains why a metric helps gain insights into the factors, and the assumption or other information useful in evaluating an OSS. It helps to identify the needed data to extract from the available portals and to automate the OSS analysis and assessment process. With the explanation of why the metrics are needed, practitioners can also better understand the factors and their assessment, which further eases the process to adopt the OSS selection models and tools in the software development practice.

\rt{Furthermore, besides the common evaluation metrics identified in this study that suits targeting any OSS, it is noticeable that the domain fitness of such targeting OSS is also of great importance. Though this study focuses on the general quality attributes of OSS, the assessment of domain fitness should be taken into account with the domain-fitting OSS candidates limited so that the evaluation effort can be largely reduced.}

Develop tools that support the research directions listed above (i.e., tools able to support and simplify the applicability of the proposed models during the evaluation  of OSS products). 

Disseminate the information that should be provided on OSS portals, so as to enable OSS producers to consider them as part of their marketing and communication strategies~\cite{DelBianco2012}.

\label{sec:Discussion}

\subsection{Threats to Validity}
We applied the structure suggested by Yin~\cite{YinCaseStudies2009} to report threats to the validity of this study and measures for mitigating them. We report internal validity, external validity, construct validity, and reliability.


% ...

\textbf{Internal Validity}. One limitation that is always a part of survey research is that surveys can only reveal the perceptions of the respondents which might not fully represent reality. However, our analysis was performed by means of semi-structured interviews, which gave the interviewers the possibility to request additional information regarding unclear or imprecise statements by the respondents. The responses were analyzed and quality-checked by a team of four researchers.

\textbf{External Validity}. Overall, a total of 23 practitioners were interviewed. We considered only experienced respondents and did not accept any interviewees with an academic background. However, we are aware that the convenience  sampling approach we adopted could be biased, even if we tried to maximize the diversity. \rd{For example, practitioners from different domains, such as those developing real-time or safety-critical systems, might have provided a different set of  answers. }
As for the projects we selected to validate the
presence of information in OSS portals, we are aware that the 100K most starred GitHub projects might not represent the whole OSS ecosystem, but we believe they might be a good representative of them. We also think that less popular projects, might only perform worst than the selected ones. 

We therefore think that threats to external validity are reasonable. However, additional responses and additional projects should be analyzed  in the future.

\textbf{Construct Validity}. The interview guidelines were developed on the basis of the previously performed surveys~\cite{DelBianco2009}\cite{Taibi2015ICSEA1}. Therefore, 
the questions are aligned with standard terminology and cover the most relevant characteristics and metrics. In addition, the survey was conducted in interviews, which allowed both the interviewees and the interviewer to ask questions if something was unclear. 

\textbf{Reliability}. The survey design, its execution, and the analysis followed a strict protocol, which allows replication of the survey. However, the open questions were analyzed qualitatively, which is always subjective to some extent, but the resulting codes were documented.

\rt{This work was based on information extracted from OSS portals and the available APIs, and therefore, reliability of the assessment depends partly on the availability and reliability of the portals. Some projects might be managed and discussed on different platforms like the different issue tracking systems, the extracted information from the available APIs might be incomplete, which may affect the assessment results. On the other hand, we identified from the interviews the most used portals in each category of the sources of information. This helps mitigate the threat to some extent.} Moreover, some projects might serve the purpose of providing resources, and not source code. \ra{However, Github API does not provide filtering functions towards excluding such projects.} We believe that, this threat could be mitigated by the large amount of projects we selected. 

\label{sec:ttv}

\section{Conclusion}
In this paper, we investigated the factors considered by companies when selecting OSS to be integrated in the software they develop, and we analyzed their availability in the OSS portals, in particular using OSS portal APIs.

We identified a set 8 factors and 74 sub-factors, together with 170  metrics that companies commonly use to evaluate and select OSS.  Unexpectedly, only a small part of the factors can be evaluated automatically, and out of 170 metrics, only 40 are available from project portals APIs.

The automated extraction of the information from the 100K most starred GitHub projects showed that only 22 metrics out of 40 returned information for all the 100K projects. 2 metrics returned information for around 80\% of the projects while another 7 for around 40\%. The other 4 metrics returned information for below 15\%.

It is important to note that the extraction consider some of the most famous OSS projects. Therefore, we can speculate that the vast majority of less common and less used projects might provide even less information. 

The result of this work enable us to create a list of updated factors and metrics,  together with the list of automatically collectable ones, that practitioners can use to select OSS.

Results can be used also by researchers to further validate the factors and metrics, or providing frameworks or tools to ease the selection of OSS. \rt{Moreover, OSS producers, and repositories might also benefit of this results to understand which information they should provide from their APIs, so as to ease the evaluation of OSS projects, and increase the adoption likelihood.}

Future work include the validation of the factors and metrics in industrial settings, reducing the subjectivity of the decisions. Moreover, we are planning to develop a tool and portal to automatically collect the information and enable the comparison of OSS projects, so as to ease the OSS selection phase. 
\label{sec:conclusion}

\section*{CRediT author statement}
\textbf{Xiaozhou Li:}  Software, Formal Analysis, Investigation, Writing - Original Draft, Visualization. \textbf{Sergio Moreschini:} Software, Formal Analysis, Investigation, Writing - Original Draft, Visualization. \textbf{Zheying Zhang:} Conceptualization, Investigation, Writing - Review and Editing. \textbf{Davide Taibi:} Conceptualization, Methodology, Funding Acquisition, Review and Editing, Supervision.





\newpage
\bibliographystyle{model1-num-names}
\bibliography{bibliography,clowee}

\section*{Appendix A: Results from the interviews}
\begin{table}[H]
\adjustbox{center}{

    \centering
    \scriptsize
    \begin{tabular}{|p{0.2cm}p{5.8cm}|p{5.4cm}|c|c|} \hline
\multicolumn{2}{|l|}{	Factor	}		&	Measure	&	\#	&	Median	\\	\hline
\multicolumn{2}{|l|}{	Community Support and Adoption	}		&		&	10	&	4.5	\\	\hline
		&	Popularity	&		&	9	&	3	\\	
		&		&	\# Watch	&	4	&	3	\\	
		&		&	\# Stars	&	13	&	3	\\	
		&		&	\# Fork	&	4	&	3	\\	
		&		&	\# Downloads	&	13	&	3	\\	
											
		&	Community reputation	&		&	11	&	3	\\	
		&		&	Member of a foundation 	&	1	&	4	\\	
		&		&	Complete administration mechanism	&	1	&	5	\\

		&		&	Fast response to issues	&	1	&	5	\\	
											
		&	Community size	&		&	13	&	3	\\	
		&		&	\# Contributors	&	11	&	4	\\	
		&		&	\# Subscribers	&	3	&	3	\\	
		&		&	Community age	&	12	&	3	\\	
		&		&	\# Involved developers per company	&	3	&	3	\\	
		&		&	\# Independent developers	&	3	&	3	\\

											
		&		&	Activeness	&	3	&	3	\\	
		&		&	Responsiveness	&	2	&	3.5	\\	
		&	Communication	&		&	6	&	3.5	\\	
		&		&	Availability of questions/answers	&	11	&	3	\\	
		&		&	Availability of forum	&	4	&	2.5	\\	
		&		&	\# Mailing lists	&	3	&	3	\\	
		&		&	Traffic on the mailing list	&	3	&	3	\\	
		&		&	Responsiveness of postings	&	4	&	4	\\	
		&		&	Friendliness	&	6	&	2.5	\\	
											
		&		&	Quality of postings	&	3	&	3	\\	
											
		&	Involvement	&		&	9	&	3	\\	
											
		&		&	Clear project management	&	1	&	5	\\

		&	Sustainability	&		&	11	&	3	\\	
		&		&	Existence of maintainer	&	11	&	3	\\

		&	Product Team	&		&	5	&	3	\\	
		&		&	Developer quality	&	3	&	4	\\	
		&		&	Developer Productivity	&	2	&	3	\\

		&	Responsiveness	&		&	1	&	5	\\	
											
		&	Scheduled updates	&		&	1	&	2.5	\\	
											
		&	Fast respond to user's needs	&		&	1	&	2.5	\\

											\hline
\multicolumn{2}{|l|}{	Documentation	}		&		&	14	&	4	\\	\hline
											
		&	Avail. of documentation/books/online docs	&		&	5	&	3	\\	
		&		&	Avg time to implement new issues	&	1	&	4	\\	
											
		&		&	Avail. of updated documentation	&	9	&	4	\\	
		&	Avail. of development process documentation	&		&	4	&	3	\\

		&		&	Avail. of getting started tutorial	&	1	&	5	\\	
											
		&	Avail. of Tutorial or Examples	&		&	5	&	5	\\

		&	Usage documentation	&		&	4	&	4	\\

		&	Avail. of best practices	&		&	4	&	4	\\

		&		&		&	4	&	3	\\

		&	Software requirements	&		&	11	&	3	\\	
		&		&	Complete doc. on SW requirements	&	1	&	5	\\	
		&	Hardware requirements	&		&	8	&	3.5	\\	
		&		&	Complete doc. on HW requirements	&	1	&	5	\\	
											
		&	Roadmap	&		&	7	&	3	\\	
		&	Test case documentation	&		&	4	&	3	\\

					\hline

	        \end{tabular}
	        }
    \label{tab:Others-1}
\end{table}										
\begin{table}[H]
\adjustbox{center}{

    \centering
    \scriptsize
    \begin{tabular}{|p{0.2cm}p{5.8cm}|p{5.4cm}|l|l|}	\hline		
    \multicolumn{2}{|l|}{	\textbf{Factor}	}	&	\textbf{Measure}	&	\textbf{\#}	&	\textbf{Median}	\\ \hline
\multicolumn{2}{|l|}{	License	}		&		&	21	&	4	\\	\hline
		&		&	License type	&	20	&	5	\\	
		&		&	Law conformance	&	9	&	5	\\	
											
		&		&	License Compatibility	&	10	&	3	\\	
		&		&	OSS obligation fullfilment 	&	1	&	5	\\	
		&		&	Existance of malicious OS obligation 	&	1	&	5	\\	
		&		&	Contagioiusness	&	1	&	5	\\	
											
		&		&	Multiple license option	&	1	&	2.5	\\

		&		&	Dual License with limited features	&	7	&	4	\\	\hline
\multicolumn{2}{|l|}{	Operational SW Characteristics	}		&		&	6	&	4	\\	\hline
		&	Trialability	&		&	5	&	3	\\	
		&		&	Available for independent verification and compile	&	1	&	5	\\	
		&		&	Provide demo for quick evaluation	&	1	&	4	\\	
		&	Independence from other SW	&		&	11	&	3	\\	
											
		&		&	Run independently	&	1	&	5	\\	
		&		&	Supports independent libraries	&	1	&	5	\\	
		&		&	Fewer dependences 	&	7	&	4	\\	
		&	Development language	&		&	5	&	4	\\	
		&		&	Mainstream dev Lang	&	4	&	4	\\	
		&		&	Language know in the company	&	2	&	4.5	\\	
		&		&	Programming language uniformity	&	5	&	4	\\	
		&	Multiplatform support	&		&	5	&	3	\\	
											
		&	Standard compliance	&		&	5	&	4	\\

		&		&	Testability	&	6	&	3.5	\\

											\hline
\multicolumn{2}{|l|}{	Maturity	}		&		&	6	&	3.5	\\	\hline
		&		&	\# forks	&	3	&	3	\\	
		&		&	Stability	&	7	&	5	\\	
		&		&	Release version stability	&	1	&	5	\\	
		&		&	\# releases 	&	10	&	4	\\	
		&		&	Release frequency	&	7	&	4	\\	
		&		&	\# releases	&	3	&	4	\\	
											
		&		&	Age (\#Years)	&	4	&	4	\\	
		&		&	\# commits	&	3	&	3	\\	
											
		&		&	Development versions	&	6	&	4	\\	
		&		&	System growth	&	9	&	4	\\	
		&		&	New feature integration	&	1	&	5	\\	
											\hline
										

\multicolumn{2}{|l|}{	Risk (Perceived risks)	}		&		&	7	&	4.5	\\	\hline
											
		&	Perceived lack of integrity	&		&	5	&	3	\\	
		&	Perceived high availability	&		&	5	&	4	\\	
		&		&	Test according to context	&	1	&	4	\\	
		&		&	Analysis and pre-examination	&	1	&	5	\\	
		&		&	Comply with business requirements	&	1	&	5	\\

		&	Strategic risks	&		&	5	&	3	\\	
											
		&		&	Influence of operation specified	&	1	&	4	\\	
											
		&	Hazard risks	&		&	5	&	4	\\	
		&		&	consequences specified	&	1	&	5	\\	
											
		&		&	Code security	&	1	&	4	\\	
		&		&	Virus scanning	&	1	&	4	\\	
											
		&		&	Risk of no mainteinance	&	1	&	2.5	\\

							\hline				        \end{tabular}
							}
    \label{tab:Others-2}
\end{table}

\begin{table}[]
\adjustbox{center}{

    \centering
    \scriptsize
    \begin{tabular}{|p{0.2cm}p{5.8cm}|p{5.4cm}|l|l|} \hline
    \multicolumn{2}{|l|}{	\textbf{Factor}	}	&	\textbf{Measure}	&	\textbf{\#}	&	\textbf{Median}	\\ \hline
\multicolumn{2}{|l|}{	Quality	}		&		&	6	&	3.5	\\	\hline
											
		&	Reliability	&		&	3	&	4	\\	
		&		&	Component reliability	&	2	&	5	\\	
		&		&	Architecture reliability	&	7	&	4	\\	
		&		&	System reliability	&	2	&	4.5	\\	
		&		&	\# Bugs (open, closed, ...)/bug density	&	8	&	4	\\	
		&		&	Average bug age	&	2	&	4.5	\\	
		&		&	Mean time between software failure (MTBF)	&	8	&	4	\\	
											
		&	Performances	&		&	4	&	4.5	\\

		&		&	Main functionality external performance standards	&	1	&	5	\\	
		&		&	Based on business	&	1	&	2.5	\\	
		&		&	Construct verification environment	&	1	&	2.5	\\	
		&		&	Comparison with similar software	&	1	&	2.5	\\	
		&	Security	&		&	15	&	4	\\	
		&		&	\# security vulnerabilities	&	12	&	3	\\	
		&		&	\#Vulnerabilities reported on the NVD portal	&	14	&	4	\\	
		&		&	Security report	&	7	&	5	\\

		&		&	Vulnerability Resolving time	&	2	&	5	\\	
		&		&	Community concern towards security	&	1	&	5	\\	
		&		&	Vulnerability impact	&	1	&	5	\\	
		&	Modularity	&		&	3	&	3	\\	
											
		&		&	Select OSS based on module	&	1	&	5	\\

		&	Portability	&		&	3	&	4	\\	
		&		&	Adaptability	&	2	&	4.5	\\	
		&		&	Installability	&	2	&	4.5	\\	
											
		&	Flexibility/Exploitability	&		&	3	&	3	\\	
											
		&		&	Support usage patterns	&	1	&	2.5	\\	
											
		&		&	Reasonable function wrapper	&	1	&	2.5	\\	
		&	Code Quality	&		&	13	&	4	\\	
		&		&	Code complexity (class, methods, ..)	&	10	&	3	\\	
		&		&	Change proneness	&	3	&	3	\\	
		&		&	Fault proneness	&	3	&	4	\\	
		&		&	Test coverage	&	4	&	4.5	\\	
		&		&	Code size	&	7	&	3	\\	
		&		&	Technical difficulty	&	3	&	4	\\

		&	Coding conventions	&		&	9	&	3	\\

		&		&	Usage of linters for checking coding conventions compliance	&	7	&	3	\\

		&	Maintainability	&		&	3	&	4	\\	
											
		&	Testability	&		&	2	&	4	\\	
		&	Changeability	&		&	2	&	3.5	\\

		&	Update/Upgrade/Add-ons/Plugin	&		&	3	&	4	\\	
											
		&		&	Update capability between versions	&	1	&	2.5	\\	
		&		&	Easy to update to new version	&	5	&	3	\\	
		&		&	API compatibility between versions	&	1	&	2.5	\\

		&	Architectural quality	&		&	5	&	3	\\

											 \hline										
        \end{tabular}
        }
    \label{tab:Others-3}
\end{table}
										 
\begin{table}[]
\adjustbox{center}{

    \centering
    \scriptsize
    \begin{tabular}{|p{0.2cm}p{5.8cm}|p{5.4cm}|l|l|} \hline
    \multicolumn{2}{|l|}{	\textbf{Factor}	}	&	\textbf{Measure}	&	\textbf{\#}	&	\textbf{Median}	\\ \hline
\multicolumn{2}{|l|}{	Trustworthiness	}		&		&	6	&	4	\\	\hline
		&	Component	&		&	4	&	3.5	\\

		&		&	Functionality	&	1	&	5	\\	
											
		&	Architecture	&		&	4	&	3	\\

		&		&	Difference with reality	&	1	&	4	\\	
		&	System	&		&	4	&	3.5	\\

		&		&	Percentage of system failure	&	1	&	5	\\

		&	OSS provider reputation	&		&	4	&	3.5	\\	
											
		&	Existence of benchmark/test	&		&	4	&	3.5	\\	
											
		&		&	Fast responsiveness to malicious affairs	&	1	&	5	\\	
		&		&	Transparency	&	1	&	4	\\

		&		&	Test cases availability	&	1	&	4	\\	
											
		&	Collaboration with other product	&		&	4	&	2.5	\\

		&		&	Even distribution among code submitters	&	1	&	2.5	\\	
		&		&	Consistent release updates pace	&	1	&	2.5	\\	
		&		&	In-time vulnerability publishing	&	1	&	2.5	\\	
		&		&	Measure-realted information (i.e. measure possibility)	&	1	&	2.5	\\

		&	Assessment results from 3rd parties	&		&	2	&	3.75	\\	
						\hline					
										
        \end{tabular}
        }
    \label{tab:Others-4}
\end{table}
\label{sec:questionnaire}
%
%
%
%






\end{document}


\begin{frontmatter}
\title{ Questionnaire on Opens Source Software (OSS) Adoption}

\author[TUNI]{Xiaozhou Li*}
\ead{xiaozhou.li@tuni.fi}

\author[TUNI]{Sergio Moreschini*}
\ead{sergio.moreschini@tuni.fi}

\author[TUNI]{Zheying Zhang}
\ead{zheying.zhang@tuni.fi}

\author[TUNI]{Davide Taibi}
\ead{davide.taibi@tuni.fi}

\address[TUNI]{Tampere University, Tampere (Finland) \\
$*$ the two authors equally contributed to the paper}

%


\begin{keyword}
Open Source \sep Software Selection \sep Open Source Adoption
\end{keyword}

\end{frontmatter}

\begin{enumerate}
    \item \textbf{Personal Information}
    \begin{enumerate}
        \item Name
        \item Age
        \item Role
        \item Time in the company (years)
        \item Unit/Department
    \end{enumerate}
    \item \textbf{General Questions regarding OSS Adoption}
    \begin{enumerate}
        \item How experienced are you with OSS?
        \item Are you following any specific model for selecting OSS or are you adapting any model to your needs? If yes, which changes are you applying to the standard model?
        \item What are the key factors that you consider when you adopt OSS? How important is each factor from 0 to 5?, where 0 means not important at all, and 5 means extremely important?
        \item What factors do you think personally important? How important is each factor from 0 to 5, where 0 means not important at all, and 5 means extremely important?
        \begin{enumerate}[label*=\arabic*.]
            \item [F1:] Community \& Support
            \item [F2:] Documentation
            \item [F3:] Economic
            \item [F4:] License
            \item [F5:] Operational SW Characteristics
            \item [F6:] Maturity
            \item [F7:] Quality
            \item [F8:] Risk
            \item [F9:] Trustworthiness
            \item [F*:] Other factors
        \end{enumerate}
        \item \textit{Note for the interviewer: }For each factor evaluated with importance higher or equal than 3, ask the interviewee to rank the sub-factors and to specify which measure they adopted to evaluate them, and from which source they commonly obtain the measures (e.g. GitHub web interface, GitHub APIs, SonarCloud, manual measurement, ...)          
    \end{enumerate}

\end{enumerate}
\begin{table}[]
    \centering
    \scriptsize
    \begin{tabular}{|p{0.2cm}p{5.6cm}|p{5.6cm}|l|} \hline
\multicolumn{2}{|l|}{	Factor	}		&	Measure	&	Vote (1-5)	\\	 \hline
\multicolumn{2}{|l|}{	Community Support and Adoption	}		&		&		\\	\hline
		&	Popularity	&		&		\\
		&		&	\# Watch	&		\\
		&		&	\# Stars	&		\\
		&		&	\# Fork	&		\\
		&		&	\# Downloads	&		\\
		&		&	Other (Please specify)	&		\\
		&	Community reputation	&		&		\\
		&		&	Member of a foundation (e.g. Apache, Linux)	&		\\
		&		&	Complete administration mechanism	&		\\
		&		&	Proactive checking on OSS quality	&		\\
		&		&	Transparent OSS quality	&		\\
		&		&	Fast response to issues, 	&		\\
				&		&	Other (Please specify)	&		\\
		&	Community creativity	&		&		\\
				&		&	Other (Please specify)	&		\\
		&	Community size	&		&		\\
		&		&	\# Contributors	&		\\
		&		&	\# Subscribers	&		\\
		&		&	Community age	&		\\
		&		&	\# Involved developers per company	&		\\
		&		&	\# Independent developers	&		\\
		&		&	Other (Please specify)	&		\\
		&	Support and service	&		&		\\
				&		&	Other (Please specify)	&		\\
		&	Community support	&		&		\\
		&		&	Activeness	&		\\
		&		&	Responsiveness	&		\\
				&		&	Other (Please specify)	&		\\
		&	Communication	&		&		\\
		&		&	Availability of questions/answers	&		\\
		&		&	Availability of forum	&		\\
		&		&	\# Mailing lists	&		\\
		&		&	Traffic on the mailing list	&		\\
		&		&	Responsiveness of postings	&		\\
		&		&	Friendliness	&		\\
		&		&	Relation between stakeholders	&		\\
		&		&	Quality of postings	&		\\
		&		&	Other (Please specify)	&		\\
		&	Involvement	&		&		\\
		&		&	Coordination	&		\\
		&		&	Clear project management	&		\\
		&		&	\# mailing list members	&		\\
		&		&	Other (Please specify)	&		\\
		&	Sustainability	&		&		\\
		&		&	Existence of maintainer	&		\\
		&		&	Maturity Implications of Mailing List Support	&		\\
		&		&	Other (Please specify)	&		\\ \hline
		    \end{tabular}
    \label{tab:Community}
\end{table}		
\begin{table}[]
    \centering
    \scriptsize
    \begin{tabular}{|p{0.2cm}p{5.6cm}|p{5.6cm}|l|} \hline
    \multicolumn{2}{|l|}{	Factor	}		&	Measure	&	Vote (1-5)	\\	 \hline
\multicolumn{2}{|l|}{	Community Support and Adoption	}		&		&		\\	\hline
		&	Paid support	&		&		\\
		&		&	Availability of official support	&		\\
		&		&	Availability of 3rd party support	&		\\
		&		&	External service provider support	&		\\
		&		&	Quality of professional support	&		\\
		&		&	Availability of training	&		\\
		&		&	Availability of end user training	&		\\
		&		&	Other (Please specify)	&		\\
		&	Failure support or maintenance	&		&		\\
				&		&	Other (Please specify)	&		\\
		&	Product Team	&		&		\\
		&		&	Developer quality	&		\\
		&		&	Productivity	&		\\
		&		&	Other (Please specify)	&		\\
		&	Other (Please specify)	&		&		\\\hline
		&	&		&		\\\hline
				&	&		&		\\\hline
						&	&		&		\\\hline
								&	&		&		\\\hline
										&	&		&		\\\hline
	
		\multicolumn{2}{|l|}{	Economic	}		&		&		\\	\hline
		&	Cost	&		&		\\
		&		&	License	&		\\
		&		&	Support	&		\\
		&		&	Training/Learning	&		\\
		&		&	Staffing	&		\\
		&		&	Promotion	&		\\
		&		&	Ownership (TCO)	&		\\
		&		&	Return of investment (ROI)	&		\\
		&		&	Other (Please specify)	&		\\
		&	Clear project management	&		&		\\
				&		&	Other (Please specify)	&		\\

		&	Profitability	&		&		\\
				&		&	Other (Please specify)	&		\\

		&	Resources and investment	&		&		\\

					&		&	Other (Please specify)		&	\\	
						&	Other (Please specify)	&	&	\\	\hline

					&	&		&		\\\hline
							&	&		&		\\\hline
									&	&		&		\\\hline
											&	&		&		\\\hline
													&	&		&		\\\hline
						    \end{tabular}
    \label{tab:Community}
\end{table}		
\begin{table}[]
    \centering
    \scriptsize
    \begin{tabular}{|p{0.2cm}p{5.6cm}|p{5.6cm}|l|} \hline
    \multicolumn{2}{|l|}{	Factor	}		&	Measure	&	Vote (1-5)	\\	 \hline
\multicolumn{2}{|l|}{	Documentation	}		&		&		\\	\hline
&	Documentation completeness	&		&		\\
			&		&	Other (Please specify)	&		\\
		&	Availability of documentation/books/online docs	&		&		\\
		&		&	Responsiveness to issue	&		\\
		&		&	Acccuracy	&		\\
		&		&	Updated Documentation	&		\\
			&		&	Other (Please specify)	&		\\ 
		&	Availability of development process documentation	&		&		\\
		&		&	Coverness	&		\\
		&		&	Readability	&		\\
			&		&	Other (Please specify)	&		\\
		&	Reference documentation	&		&		\\
		&		&	Available Get start document	&		\\
		&		&	Reference Completeness	&		\\
			&		&	Other (Please specify)	&		\\
		&	Tutorial/Guidelines documentation	&		&		\\
		&		&	How to start from 0	&		\\
		&		&	Coverage	&		\\
		&		&	Accuracy	&		\\
	&		&	Other (Please specify)	&		\\
		&	Usage documentation	&		&		\\
		&		&	Instructions and Examples	&		\\
		&		&	Completeness	&		\\
		&		&	Readability	&		\\
			&		&	Other (Please specify)	&		\\
		&	Best practices	&		&		\\
		&		&	Classic Best example and hwo to optimize	&		\\
		&		&	Guidance	&		\\
			&		&	Other (Please specify)	&		\\
		&	Community's experience	&		&		\\
			&		&	Other (Please specify)	&		\\
		&	Availability of architectural documentation	&		&		\\
		&		&	Accuracy	&		\\
			&		&	Other (Please specify)	&		\\
		&	Distribution media	&		&		\\
			&		&	Other (Please specify)	&		\\
		&	Software requirements	&		&		\\
		&		&	Completeness	&		\\
			&		&	Other (Please specify)	&		\\
		&	Hardware requirements	&		&		\\
			&		&	Other (Please specify)	&		\\
		&	Dissemination	&		&		\\
			&		&	Other (Please specify)	&		\\
		&	Road map	&		&		\\
			&		&	Other (Please specify)	&		\\
		&	Test case documentation	&		&		\\
	&		&	Other (Please specify)	&		\\
		\hline		
		
		    \end{tabular}
    \label{tab:Community}
\end{table}		
\begin{table}[]
    \centering
    \scriptsize
    \begin{tabular}{|p{0.2cm}p{5.6cm}|p{5.6cm}|l|} \hline
\multicolumn{2}{|l|}{	Factor	}		&	Measure	&	Vote (1-5)	\\	 \hline
	\multicolumn{2}{|l|}{	Documentation	}		&		&		\\	\hline
				&	Quality report	&		&		\\
		&		&	Coverage	&		\\
		&		&	Accuracy	&		\\
			&		&	Other (Please specify)	&		\\
		&	Coding comments	&		&		\\

		&		&	Readability	&		\\
&		&	Other (Please specify)		&	\\	
						&	Other (Please specify)	&	&	\\ \hline
											&	&		&		\\\hline
																&	&		&		\\\hline
																					&	&		&		\\\hline
																										&	&		&		\\\hline
																															&	&		&		\\\hline
		\multicolumn{2}{|l|}{	License	}		&		&		\\	\hline
		&		&	License type	&		\\
		&		&	Law conformance	&		\\
		&		&	Other (Please specify)	&		\\
		
						&	Other (Please specify)	&	&	\\	\hline
											&	&		&		\\\hline
																&	&		&		\\\hline
																					&	&		&		\\\hline
																										&	&		&		\\\hline
																															&	&		&		\\\hline
\multicolumn{2}{|l|}{	Operational SW Characteristics	}		&		&		\\	\hline
		&	Trialability	&		&		\\
		&		&	Available for independent verification and compile	&		\\
		&		&	Provide demo for quick evaluation	&		\\
			&		&	Other (Please specify)		&	\\
		&	Independence from other SW	&		&		\\
		&		&	Adopted SW architecture	&		\\
		&		&	Run independently	&		\\
		&		&	Supports independent libraries	&		\\
		&		&	Fewer dependences 	&		\\
			&		&	Other (Please specify)		&	\\
		&	Development language	&		&		\\
		&		&	Mainstream dev Lang	&		\\
		&		&	Language know in the company	&		\\
		&		&	Programming language uniformity	&		\\
			&		&	Other (Please specify)		&	\\
		&	Multiplatform support	&		&		\\
		&		&	Portability	&		\\
			&		&	Other (Please specify)		&	\\
		&	Standard compliance	&		&		\\
	&		&	Other (Please specify)		&	\\
		&	Other (Please specify)	&		&		\\
		\hline
							&	&		&		\\\hline
												&	&		&		\\\hline
																	&	&		&		\\\hline
																						&	&		&		\\\hline
																											&	&		&		\\\hline

    \end{tabular}
    \label{tab:Community}
\end{table}

\begin{table}[]
    \centering
    \scriptsize
    \begin{tabular}{|p{0.2cm}p{5.6cm}|p{5.6cm}|l|} \hline
    \multicolumn{2}{|l|}{	Factor	}		&	Measure	&	Vote (1-5)	\\	 \hline
\multicolumn{2}{|l|}{	Maturity	}		&		&		\\	\hline
		&		&	\# forks	&		\\
		&		&	Stability	&		\\
		&		&	Release version stability	&		\\
		&		&	\# bugs	&		\\
		&		&	Release frequency	&		\\
		&		&	\# releases	&		\\
		&		&	\# releases per period	&		\\
		&		&	Age (\#Years)	&		\\
		&		&	\# commits	&		\\
		&		&	\# bug reports	&		\\
		&		&	Development versions	&		\\
		&		&	System growth	&		\\
		&		&	New feature integration	&		\\

									&		&	Other (Please specify)		&	\\	
						&	Other (Please specify)	&	&	\\		\hline
			&	&		&		\\\hline
				&	&		&		\\\hline
					&	&		&		\\\hline
						&	&		&		\\\hline
							&	&		&		\\\hline
		\multicolumn{2}{|l|}{	Quality	}		&		&		\\	\hline
		&	Resilience	&		&		\\
				&		&	Other (Please specify)	&		\\
		&	Reliability	&		&		\\
		&		&	Component reliability	&		\\
		&		&	Architecture reliability	&		\\
		&		&	System reliability	&		\\
		&		&	\# faults (open, closed, ...)/Fault density	&		\\
		&		&	Average fault age	&		\\
		&		&	Mean time between software failure (MTBF)	&		\\
		&		&	Other (Please specify)	&		\\
		&	Performances	&		&		\\
		&		&	Scalability	&		\\
		&		&	Other (Please specify)	&		\\
		&		&	Main functionality external performance standards	&		\\
		&		&	Based on business	&		\\
		&		&	Construct verification environment	&		\\
		&		&	Comparison with similar software	&		\\
				&		&	Other (Please specify)	&		\\
		&	Security	&		&		\\
		&		&	\# security vulnerabilities	&		\\
		&		&	\# security vulnerabilities on NVD portal	&		\\
		&		&	Information for security	&		\\
		&		&	Other (Please specify)	&		\\
		&	Modularity	&		&		\\
		&		&	Other (Please specify)	&		\\
		&	Usability	&		&		\\
		&		&	Understandability	&		\\
		&		&	Operability	&		\\
		&		&	Learnability	&		\\
		&		&	Other (Please specify)	&		\\
	\hline
		
		    \end{tabular}
    \label{tab:Community}
\end{table}

\begin{table}[]
    \centering
    \scriptsize
    \begin{tabular}{|p{0.2cm}p{5.6cm}|p{5.6cm}|l|} \hline
    \multicolumn{2}{|l|}{	Factor	}		&	Measure	&	Vote (1-5)	\\	 \hline

		\multicolumn{2}{|l|}{	Quality	}		&		&		\\	\hline
				&	Portability	&		&		\\
		&		&	Adaptability	&		\\
		&		&	Installability	&		\\
		&		&	Other (Please specify)	&		\\
		&	Flexibility/Exploitability	&		&		\\
		&		&	Other (Please specify)	&		\\
		&	Code Quality	&		&		\\
		&		&	Code complexity (class, methods, ..)	&		\\
		&		&	Change proneness	&		\\
		&		&	Fault proneness	&		\\
		&		&	Test coverage	&		\\
		&		&	Code size	&		\\
		&		&	Technical difficulty	&		\\
		&		&	Other (Please specify)	&		\\
		&	Longevity	&		&		\\
		&		&	Age of product	&		\\
		&		&	Version number	&		\\
		&		&	Other (Please specify)	&		\\
		&	Coding conventions	&		&		\\
		&		&	Method calling conventions	&		\\
		&		&	Class naming conventions	&		\\
		&		&	Loop/Switch/If conventions	&		\\
		&		&	Public/private conventions	&		\\
		&		&	Overriding converntions	&		\\
		&		&	Other (Please specify)	&		\\
		&	Coding practices	&		&		\\
		&		&	Usage of Linters	&		\\
		&		&	Definition of customs/rules for Linters	&		\\
		&		&	Issue/Commit traceability	&		\\
		&	Maintainability	&		&		\\	
			&		&	Other (Please specify)		&	\\
		&	Analyzability	&		&		\\
			&		&	Other (Please specify)		&	\\
		&	Testability	&		&		\\
			&		&	Other (Please specify)		&	\\
		&	Changeability	&		&		\\
		&		&	Availability and type of Defect management system	&		\\
		&		&	Availability and type of Version management system	&		\\
		&		&	Opened defects	&		\\
			&		&	Other (Please specify)		&	\\
		&	Source code changes/modifiability	&		&		\\
			&		&	Other (Please specify)		&	\\
		&	Resolved defects	&		&		\\
			&		&	Other (Please specify)		&	\\
		&	Fixed defects	&		&		\\
			&		&	Other (Please specify)		&	\\
		&	Enhancements	&		&		\\
			&		&	Other (Please specify)		&	\\
		&	Resolving time for non-fixed defects	&		&		\\
			&		&	Other (Please specify)		&	\\
		&	Resolving time for fixed defects	&		&		\\
			&		&	Other (Please specify)		&	\\
	\hline

    \end{tabular}
    \label{tab:Community}
\end{table}

\begin{table}[]
    \centering
    \scriptsize
        \begin{tabular}{|p{0.2cm}p{5.6cm}|p{5.6cm}|l|} \hline
    \multicolumn{2}{|l|}{	Factor	}		&	Measure	&	Vote (1-5)	\\	 \hline
	\multicolumn{2}{|l|}{	Quality	}		&		&		\\	\hline

		&	Resolved defects life cycle	&		&		\\
			&		&	Other (Please specify)		&	\\
		&	Fixed defects life cycle	&		&		\\
			&		&	Other (Please specify)		&	\\
		&	Detect initiated source code changes	&		&		\\
			&		&	Other (Please specify)		&	\\
		&	Defect proneness	&		&		\\
			&		&	Other (Please specify)		&	\\
		&	Defect Management	&		&		\\
			&		&	Other (Please specify)		&	\\
		&	Version Management	&		&		\\
		&		&	Other (Please specify)	&		\\
		&	Update/Upgrade/Add-ons/Plugin	&		&		\\
		&		&	Other (Please specify)	&		\\
		&	Visibility	&		&		\\
			&		&	Other (Please specify)		&	\\
		&	Development process	&		&		\\
			&		&	Other (Please specify)		&	\\
		&	As-is utility/Quality in use/External quality	&		&		\\
		&		&	Effectiveness	&		\\
		&		&	Efficiency	&		\\
		&		&	Other (Please specify)	&		\\
		&	Satisfaction	&		&		\\
		&		&	Functionality/Non-functionality	&		\\
		&		&	Usefulness	&		\\
		&		&	Trust	&		\\
		&		&	Pleasure	&		\\
		&		&	Comfort	&		\\
		&		&	Other (Please specify)	&		\\
		&	Freedom from risks	&		&		\\
		&		&	Economic risk migration	&		\\
		&		&	Health \& Safety risk migration	&		\\
		&		&	Environmental risk migration	&		\\
		&		&	Other (Please specify)	&		\\
		&	Context coverage	&		&		\\
		&		&	Context completeness	&		\\
		&		&	Flexibility	&		\\
		&		&	Other (Please specify)	&		\\
		&	Architecture	&		&		\\
		&		&	Dependence \& constraints	&		\\
		&		&	Other (Please specify)	&		\\
		&		&	Support for Long Term Evolution	&		\\
			&		&	Other (Please specify)		&	\\
		&	Other (Please specify)	&		&		\\ 	\hline
	&	&		&		\\\hline
		&	&		&		\\\hline
			&	&		&		\\\hline
				&	&		&		\\\hline
					&	&		&		\\\hline

    \end{tabular}
    \label{tab:Community}
\end{table}

\begin{table}[]
    \centering
    \scriptsize
        \begin{tabular}{|p{0.2cm}p{5.6cm}|p{5.6cm}|l|} \hline
    \multicolumn{2}{|l|}{	Factor	}		&	Measure	&	Vote (1-5)	\\	 \hline

\multicolumn{2}{|l|}{	Risk (Perceived risks)	}		&		&		\\	
		&	Perceived lack of confidentiality	&		&		\\
						&		&	Other (Please specify)		&	\\	
		&	Perceived lack of integrity	&		&		\\
						&		&	Other (Please specify)		&	\\	
		&	Perceived high availability	&		&		\\
		&		&	Test according to context	&		\\
		&		&	Analysis and pre-examination	&		\\
		&		&	Comply with business requirements	&		\\
						&		&	Other (Please specify)		&	\\	
		&	Perceived high structural assurance	&		&		\\
						&		&	Other (Please specify)		&	\\	
		&	Business Risks	&		&		\\
						&		&	Other (Please specify)		&	\\	
		&	Strategic risks	&		&		\\
						&		&	Other (Please specify)		&	\\	
		&	Operational risks	&		&		\\
		&		&	Influence of operation specified	&		\\
						&		&	Other (Please specify)		&	\\	
		&	Financial risks	&		&		\\
						&		&	Other (Please specify)		&	\\	
		&	Hazard risks	&		&		\\
		&		&	Consequences specified	&		\\
		&		&	Moment metrics	&		\\
		&		&	Code security	&		\\
		&		&	Virus scanning	&		\\
		&		&	Other (Please specify)	&		\\
										&	Other (Please specify)	&	&	\\
										\hline
											&	&		&		\\\hline
												&	&		&		\\\hline
													&	&		&		\\\hline
														&	&		&		\\\hline
															&	&		&		\\\hline
\multicolumn{2}{|l|}{	Trustworthiness	}		&		&		\\	\hline
		&	Component	&		&		\\
		&		&	Component development \& certificate processes	&		\\
		&		&	Component reputation \& user community	&		\\
		&		&	Functionality	&		\\
		&		&	Other (Please specify)	&		\\
		&	Architecture	&		&		\\
		&		&	Dependence \& constraints	&		\\
		&		&	Other (Please specify)	&		\\

		&	System	&		&		\\
		&		&	Component installation	&		\\
		&		&	Delivery	&		\\
		&		&	Other (Please specify)	&		\\

		&	Functional requirements satisfaction	&		&		\\
		&		&	Percentage of functional requirements implemented	&		\\
		&		&	Compliant with description	&		\\
		&		&	Low extra useless features	&		\\
		&		&	# companies adopting it	&		\\
		&		&	Performance and Reliability	&		\\
				&		&	Other (Please specify)	&		\\
		 \hline
\end{tabular}
    \label{tab:Community}
\end{table}

\begin{table}[]
    \centering
    \scriptsize
        \begin{tabular}{|p{0.2cm}p{5.6cm}|p{5.6cm}|l|} \hline
    \multicolumn{2}{|l|}{	Factor	}		&	Measure	&	Vote (1-5)	\\	 \hline
    \multicolumn{2}{|l|}{	Trustworthiness	}		&		&		\\	\hline
    &	Customer satisfaction/requirements	&		&		\\
		&		&	Matching customer satisfaction/Features	&		\\
		&		&	Function analysis and verification evaluation	&		\\
				&		&	Other (Please specify)	&		\\
		&	Interoperability	&		&		\\
				&		&	Other (Please specify)	&		\\
		&	Facilities for modifying and customizing	&		&		\\
				&		&	Other (Please specify)	&		\\
		&	OSS provider reputation	&		&		\\
				&		&	Other (Please specify)	&		\\
		&	Interface language localization	&		&		\\
				&		&	Other (Please specify)	&		\\
		&	Existence of benchmark/test	&		&		\\
		&		&	Fast responsiveness to Issues	&		\\
		&		&	Fast responsiveness to malicious affairs	&		\\
		&		&	Transparency	&		\\
		&		&	Fame of community	&		\\
		&		&	Support for mainstrame languages	&		\\
		&		&	Test cases availability	&		\\
		&		&	Basic functions covering	&		\\
				&		&	Other (Please specify)	&		\\
		&	Collaboration with other product	&		&		\\
		&		&	Commercial vendor integrations	&		\\
		&		&	Community-created integrations	&		\\
		&		&	The percentage of existing needed integrations	&		\\
		&		&	Other (Please specify)	&		\\

										&	Other (Please specify)	&	&	\\ \hline
										&	&		&		\\\hline
										&	&		&		\\\hline
										&	&		&		\\\hline
										&	&		&		\\\hline
										&	&		&		\\\hline
										\multicolumn{2}{|l|}{	Others	}		&		&		\\ \hline
		&	Staffing	&		&		\\
		&		&	Productivity	&		\\
		&		&	Salary of employees	&		\\
		&		&	Other (Please specify)	&		\\
		&	Technical environment	&		&		\\
		&	Assessment results from 3rd parties	&		&		\\
		&	Voluntariness of use	&		&		\\
		&	Results demonstrability	&		&		\\
		&	Configuration management	&		&		\\
		&	Personal interest	&		&		\\
		&	Regulation and political influence	&		&		\\
		&	Accomplishment	&		&		\\
		&	Ethical Reasons	&		&		\\
		&	Experience stimulation	&		&		\\
		&	Other (Please specify)	&		&		\\ \hline
		&	&		&		\\\hline
		&	&		&		\\\hline
		&	&		&		\\\hline
		&	&		&		\\\hline
		&	&		&		\\\hline
\end{tabular}
    \label{tab:Community}
\end{table}